\documentclass{emulateapj}
\usepackage[utf8]{inputenc}
\usepackage[titletoc,title]{appendix}

\newcommand{\be}{\begin{equation}}
\newcommand{\ee}{\end{equation}}

\usepackage{xcolor}
\usepackage{hyperref}
\usepackage{breakurl}
\usepackage{amsmath}
\usepackage{graphicx}
\usepackage{verbatim}
\usepackage{booktabs}

\shorttitle{Planet Nine Stability Survey}
\shortauthors{Becker et al. }

\begin{document}
\title{Evaluating the Dynamical Stability of Outer Solar System Objects \\
in the Presence of Planet Nine}
\author{
Juliette C. Becker\altaffilmark{1},
Fred C. Adams\altaffilmark{1,2},
Tali Khain\altaffilmark{2}, Stephanie J. Hamilton\altaffilmark{2}, David Gerdes\altaffilmark{1,2}}

\email{jcbecker@umich.edu}
\altaffiltext{1}{Astronomy Department, University of Michigan, Ann Arbor, MI 48109, USA}
\altaffiltext{2}{Physics Department, University of 
Michigan, Ann Arbor, MI 48109, USA}


\begin{abstract}
We evaluate the dynamical stability of a selection of outer solar system objects in the presence of the proposed new Solar System member Planet Nine. We use a Monte Carlo suite of numerical N-body integrations to construct a variety of orbital elements of the new planet and evaluate the dynamical stability of eight Trans-Neptunian objects (TNOs) in the presence of Planet Nine. These simulations show that some combinations of orbital elements ($a,e$) result in Planet Nine acting as a stabilizing influence on the TNOs, which can otherwise be destabilized by interactions with Neptune. These simulations also suggest that some TNOs transition between several different mean-motion resonances during their lifetimes while still retaining approximate apsidal anti-alignment with Planet Nine. This behavior suggests that remaining in one particular orbit is not a requirement for orbital stability. As one product of our simulations, we present an {\it a posteriori} probability distribution for the semi-major axis and eccentricity of the proposed Planet Nine based on TNO stability. This result thus provides additional evidence that supports the existence of this proposed planet. We also predict that TNOs can be grouped into multiple populations of objects that interact with Planet Nine in different ways: one population may contain objects like Sedna and 2012 VP$_{113}$, which do not migrate significantly in semi-major axis in the presence of Planet Nine and tend to stay in the same resonance; another population may contain objects like 2007 TG$_{422}$ and 2013 RF$_{98}$, which may both migrate and transition between different resonances. \end{abstract}

\maketitle

\section{Introduction}
\label{sec:intro}

In our solar system, a large population of small, rocky objects resides beyond the orbit of Neptune, and the collective structure of this population is anomalous, exhibiting trends unexplained by random chance. Many of these objects appear to occupy a region close to the plane containing the eight known planets, leading to this region being called the Kuiper Belt. The existence of these objects has implications for the formation mechanism of our solar system; however, we have yet discovered only a small fraction of the objects orbiting beyond Neptune.
Since the turn of the century, many new objects have been discovered in the Kuiper Belt. The subset of objects orbiting outside of Neptune's orbit are called Trans-Neptunian objects (TNOs), and they often have dynamically interesting orbits. In particular, some of these objects have large semi-major axes and large perihelion distances, including, for example, Sedna \citep{sedna}, 2004 VN$_{112}$ \citep{2004vn112}, 2010 GB$_{174}$ \citep{gb1074}, and 2012 VP$_{113}$ \citep{st14}.

When \citet{st14} reported the discovery of 2012 VP$_{113}$, they also noted a curious clustering in argument of perihelion for the population of TNOs with high-\emph{a}, high-\emph{q} orbits. The authors proposed that the high-\emph{q} orbits could be generated in three ways: first, by the ejection of a solar system body that left behind the observed clustering as a signature of its ejection; second, through a stellar fly-by encounter that perturbed the orbits of some TNOs into their current configurations \citep{star1}, where such interactions are relatively common in the birth cluster \citep{star2}; third, through the presence of an additional planet in the solar system. Notably, this third mechanism could also explain the clustering of this population of TNOs, as the proposed ninth planet's repeated secular interactions with the shorter-period TNOs could force the TNOs to keep their $\omega$ confined to be near either 0 degrees or 180 degrees (for example, this could occur via the Kozai mechanism).

\citet{bb16} also suggested the existence of an additional planet (Planet Nine), which is differentiated from the potential planet discussed in \citet{st14} in the way it interacts with the TNOs. The \citet{bb16} version of Planet Nine functions by explaining both the apsidal and ascending node alignment of a selection of objects in the Kuiper Belt. The objects under consideration in \citet{bb16} were two overlapping sets of objects: first, those that have perihelion distance $q>30$ AU and semi-major axis $a>150$ AU, while being dynamically stable in the presence of Neptune; second, any objects which have $q>30$ AU and semi-major axis $a>250$ AU, all of which exhibit clustering in $\varpi$ (when $\varpi$ is defined as the longitude of perihelion, $\varpi = \omega + \Omega$). The orbit for Planet Nine presented in \citet{bb16} was a rough estimate, with semi-major axis $a=700$ AU, eccentricity $e=0.6$, inclination $i=30$ degrees, longitude of ascending node $\Omega$ = 113 degrees, and argument of perihelion $\omega = 150$ degrees. The authors estimated the planet to have a mass of roughly 10 Earth masses, but all of these orbit predictions were noted as approximate. This mass estimate was supported in a follow-up effort by the predicting authors \citep{bb162}, which placed constraints on the orbital elements of the potential Planet Nine by using N-body simulations to determine which Planet Nine realizations lead to clustered TNOs in simulations. In this work, the authors determined that a 10 Earth mass Planet Nine was more likely to recreate the observed clustering than a 20 Earth mass Planet Nine. 

The possibility of a new planet has led to a great deal of recent work, e.g., to evaluate the likelihood of Planet Nine's existence given the known properties of the Solar System. For example, \citet{cassini} and \citet{cassini2} examined the measured Earth-Saturn distance, and determined that Planet Nine is likely to have a true anomaly near 117 degrees (based on the upper limit to the amplitude of the residuals for that distance measurement). In complementary work, \citet{renu} evaluated the potential resonant behavior that Planet Nine could invoke in the population of long-period TNOs, and predicted that Planet Nine should have a semi-major axis of $a=665$ AU in order to support stability-boosting resonances. Other authors 
\citep{lawler,dlfdlf1} have examined the dynamical effects that Planet Nine would have on the populations of objects that exist in the Solar System. Finally, additional work was carried out to explain how Planet Nine would fit into our existing picture of the Solar System. For example, multiple groups found that the existence of such a ninth planet can be invoked to explain the six-degree obliquity of the sun \citep{bailey, lai, gomes}. Several previous works have also suggested that the mechanism by which Planet Nine shapes the orbits of the TNOs may be orbital resonances \citep{bb16, beust, renu, sarah}.

The conclusion in \citet{bb16} was based only on six of the (at the time) thirteen discovered extreme TNOs. These six TNOs (2004 VN$_{112}$, 2007 TG$_{422}$, 2010 GB$_{174}$, 2012 VP$_{113}$, 2013 RF$_{98}$, and Sedna) were chosen because they exhibit clustering in $\varpi$ and have $a>250$ AU, meaning they are expected to be influenced by Planet Nine. \citet{ST_tnos} also announced the discovery of two new objects (2014 SR$_{349}$ and 2013 FT$_{28}$) that also fit into the class of objects used by \citet{bb16} to predict the existence of Planet Nine.
In order for Planet Nine to be capable of forcing the orbits of the TNOs into aligned configurations, the TNOs must remain dynamically stable on secular timescales, to allow apsidal alignment to occur. Objects whose orbital elements are affected by short-period scattering events can be used to understand the scattering event itself, but not the long-period, secular effects in the system. 
Numerical simulations in \citet{bb16} and \citet{ST_tnos} determined that some of the high semi-major axis ($a>250$) objects' orbits change significantly over 1 Gyr timescales. 
They used numerical simulations to evaluate the stability of these bodies in the presence of the four giant planets, and found that some objects were not dynamically stable. If two of the six objects used to infer the existence of Planet Nine are dynamically unstable (for example, being susceptible to scattering events) in the presence of Neptune, it is less likely that those objects could reside in their current orbits long enough to be influenced by Planet Nine and become apsidally aligned via that mechanism. \citet{dlfdlf1} also found that the six aforementioned extreme TNOs can become dynamically unstable on relatively short timescales in the presence of both Neptune and the nominal Planet Nine, which could potentially prevent the observed apsidal alignment from occurring on secular timescales. 

Of course, dynamical stability is a function of timescale. The objects with perihelion distances in the range 30-40 AU can be termed a part of the scattered disk \citep{scat_disk}. The objects in this population are characterized by repeated (potentially scattering) interactions with Neptune \citep{nesvorny}. An integration of the solar system run for an indefinite amount of time (without the presence of Planet Nine) will eventually lead to all objects in the scattered disk leaving the solar system. However, our solar system is only 4.5 Gyr old, and it does not take this long for Planet Nine to align the TNOs into the pattern reported in \citet{bb16}. For this reason, `dynamical stability' in this work refers to objects which remain in orbits comparable to their current orbits for 4.5 Gyr. 

\citet{ST_tnos} reported two new TNOs in the regime of interest for consideration of Planet Nine, and the population of discovered TNOs and Kuiper Belt Objects is rapidly growing. The Canada-France Ecliptic Plane Survey \citep{cfeps}, which ran from 2003--2007, detected 169 TNOs with a preference for larger TNOs (size $R>$ 100 km). A 32-square-degree survey running from 2011--2012 by \citet{alex2} detected 77 TNOs. The Outer Solar System Origins Survey \citep{osss}, which is currently in progress, has thus far detected 85 TNOs. The Panoramic Survey Telescope and Rapid Response System (Pan-STARRS) is intended to discover comets and asteroids (particularly near-Earth objects), but has also found Kuiper Belt Objects \citep{niku} and giant planet Trojans \citep{trojans1, trojans2, trojans3}. In addition to these dedicated Solar System searches, cosmological searches also allow for the serendipitous discovery of foreground TNOs. For example, the Dark Energy Survey has experienced great success in discovering TNOs and other solar system objects \citep{gerdes1, des_survey}, including a new dwarf planet \citep{gerdes2}.

This ever-growing population of TNOs will allow for increasingly stronger constraints on the possible orbital elements of Planet Nine and its current location in the Solar System. These objects will also provide a clearer picture of the dynamical regimes of bodies in the outer Solar System, where these orbits are sculpted by Neptune on the inside and could be influenced by Planet Nine from the outside. 

In this paper, we use as our sample the TNOs with $a>250$ AU from \citealt{bb16} and two newly discovered objects \citep[announced in][]{ST_tnos} expected to be dynamically stable in the presence of the solar system objects. We use N-body techniques to simulate the behavior of these objects in the presence of Planet Nine to place limits on the possible orbital elements ($a,e$) of Planet Nine. 
In Section \ref{sec:num_methods}, we describe the sets of simulations carried out in this work and present some results. Our treatment uses a Monte Carlo sample of 1500 Planet Nine realizations, and thus extends most previous work, which generally considered only a single nominal orbit or small number (generally $N\le3$) of potential orbits. The simulations enable us to estimate the mean lifetime of each TNO in the presence of each of the Planet Nine realizations under consideration. In Section \ref{sec:bayes}, we develop and provide a posterior probability distribution for the most likely values ($a,e$) for the Planet Nine orbit, given that we observe the eight TNOs in their current orbits. In Section \ref{sec:neptune}, we discuss the different dynamical instability mechanisms contributing to the potentially shortened lifetimes for the TNOs, and identify some interesting differences between objects in our sample. In Section \ref{sec:resonance}, we explore the relationship between the period of our injected Planet Nine and the period of each TNO, and show that TNOs living in dynamically stable configurations tend to attain integer period ratios with Planet Nine. Remarkably, some TNOs do not remain in the same commensurability for their entire lifetimes; instead, they transition between multiple near-resonant locations. The paper concludes, in Section \ref{sec:conclusion}, with a summary of our main results and a discussion of avenues for future study.  


\section{Numerical Simulations of TNO orbital evolution in the presence of Planet Nine}
\label{sec:num_methods}
\citet{bb162} provided a relative posterior probability distribution for the ($a,e$) of Planet Nine. This posterior was constructed using clustering arguments of the same type that were used in \citet{bb16} to predict the existence of Planet Nine. The logic used in both these works can be summarized as follows: the observed TNOs' orbits are aligned in physical space (the longitudes of perihelion $\varpi$ and longitudes of ascending node $\Omega$ are confined to a narrow range in angles instead of uniformly populating all allowed angles), and the probability of this occurring by chance is low, even accounting for potential bias in the observations. Numerical N-body simulations of a selection of test-particles at varying orbital radii do not recreate the observed clustering unless Planet Nine is included in the simulations; even then, some orbits of Planet Nine are more likely to recreate the observed clustering than others. From a suite of numerical integrations, \citet{bb162} predicted the combinations of ($a,e$) that were most likely to allow a population of test particles that exhibited the observed clustering in longitude of perihelion.  

The physical orientation of orbits is not the only observed physical property that can be measured from the known TNOs. The most basic property that the observed TNOs share is orbital stability: although merely seeing the TNO orbits today does not ensure that they are dynamically stable in those current orbits, the observed \citep{st14, bb16} physical clustering of the orbit directions suggests that the orbits have been dynamically stable for a long enough time for this alignment to develop. Whether or not the aligning agent is the theorized Planet Nine, this alignment would have to have taken place over secular timescales, suggesting that these TNOs must have been dynamically stable over such timescales. 

The dynamical stability of these objects is thus a fundamental and necessary property. As such, any allowable Planet Nine would have to allow the continued dynamical stability of the TNOs over secular and solar system lifetimes. 
To evaluate the likelihood of any particular realization of Planet Nine, we must evaluate the lifetimes of the TNOs in the presence of said Planet Nine.

\subsection{Numerical Methods}
To determine how the lifetimes of the TNOs vary in the presence of different combinations of the semi-major axis and eccentricity of Planet Nine, we run a large number of numerical N-body integrations, each including one potential realization of Planet Nine and the population of TNOs we are testing. This population includes the six TNOs considered in \citet{bb16} (2004 VN$_{112}$,
2007 TG$_{422}$, 2010 GB$_{174}$, 2012 VP$_{113}$, 2013 RF$_{98}$, and Sedna) and two additional objects discovered by \citet{ST_tnos} (2014 SR$_{349}$ and 2013 FT$_{28}$) that appear to fit in the same dynamical class as the previous six. The orbital elements of all eight objects in our sample are reported in Table \ref{tab:stab}. The objects we consider in this work are those with semi-major axis $a > 250$ AU and perihelion distance $q > 30$ AU. This is a different sample, with a stricter semi-major axis cut, than was originally used in \citet{st14} to identify the clustering effect \citep[][used $a > 150$ AU]{st14}. We limited the objects considered in this paper to those with $a>250$ AU for three reasons: first, these are the objects which are found in \citet{bb16} to exhibit both confinement in the longitude of perihelion and in the longitude of ascending node; second, these large-$a$ objects exhibit a variability in semi-major axis \citep{bb16} that can potentially lead to dynamical instability; third, we expect the dynamics of these long period orbits to be dominated by Planet Nine rather than by Neptune \citep{ST_tnos}.

We choose to explore the effect of Planet Nine's $a$ and $e$ because the orbital angles of Planet Nine are directly oppositional to those of (all but one of) the discovered ETNOs, and the current location of Planet Nine can be estimated using other methods \citep[for example,][]{cassini, cassini2}. 
For this work, the inclination and orbital angles of Planet Nine were taken to be the nominal values presented in \citet{bb16}. The robust consideration of a wider range of angles is beyond the scope of this work, but we report a basic reproduction of the experiments of this work while allowing the orbital angles to vary in Appendix \ref{app2}. We choose to fix these angles for the remainder of the main manuscript for two reasons: first, to examine the effect of ($a,e$), we take a cross section in the other angles in order to remove the degeneracies that would be imposed by allowing these angles to vary; second, considering all orbital angles would be too computationally intensive, so we choose the best values we can and proceed under the assumption that altering these angles will not change the broad trends of the results. 

To evaluate the dynamical stability (and thus, expected mean orbital lifetimes) of these eight TNOs in the presence of Planet Nine, we ran 1500 numerical N-body integrations using the hybrid symplectic and Bulirsch-Stoer (B-S) integrator built into \texttt{Mercury6} \citep{m6}, and conserved energy to 1 part in $10^{10}$.
We replaced the three inner giant planets (Jupiter, Saturn and Uranus) with a solar J2 \citep[as done in][]{bb16}. We included Neptune as an active particle because the orbital motion of Neptune may induce scattering or resonant effects on the TNOs, potentially leading to dynamical instabilities (rapid, drastic orbital evolution for the TNOs). For each realization, we included each TNO with orbital elements drawn from observational constraints, sampling each orbital element from reported 1$\sigma$ errors. Then, we integrated each of the 1500 realizations forward for 4.5 Gyr using computational resources provided by \citet{xsede}. For data management, we used the \texttt{pandas} python package \citep{pandas}. 

Some of the TNOs (2013 RF$_{98}$, for example) have very rough observational priors. 
Since we are sampling from observational priors in the orbital elements we assign to these small bodies, the lifetimes of these objects (compared to Sedna, whose orbital properties are much better constrained) include the degeneracy of both Planet Nine and also of the TNO itself. This leads to smoothed distributions, which are effectively convolutions of the true lifetimes with an uncertainty kernel including the errors of the measured orbital properties of the object. 

\begin{figure} 
   \centering
   \includegraphics[width=3.2in]{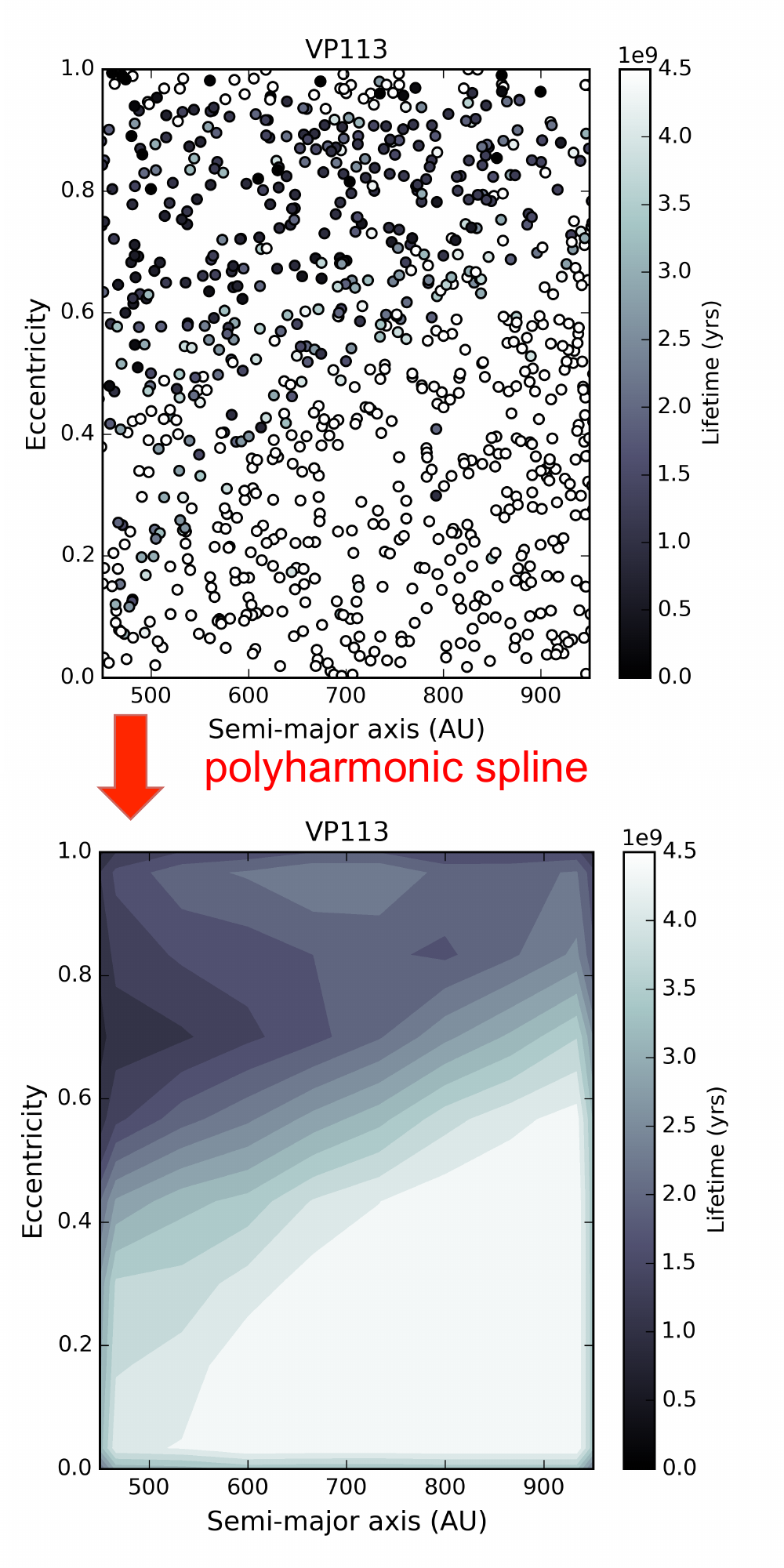} 
   \caption{(Top panel) The points for one object in our sample, 2012 VP$_{113}$, which are color coded by the amount of time the system remained dynamically stable. (Bottom panel) The same points, turned into a contour plot by use of a polyharmonic spline. The contour plot allows for easier comparison of the posterior for each object, and visualizes the interpolation  up for the Monte Carlo sampling.  }
   \label{fig:spline}
\end{figure}

The final result of these simulations is 1500 measures of dynamical lifetime for each object, which can be plotted on a ($a, e$) grid as shown in the top panel of Figure \ref{fig:spline} (which shows the points for 2012 VP$_{113}$). We do not expect the lifetimes to form a smooth function for several reasons: first, for each object, we draw orbital parameters from observational priors, resulting in some expected scatter in results for even a single Planet Nine realization; second, chaotic effects will cause scatter in outcomes between realizations. For that reason, we must run enough simulations that we can treat the averaged value from all results near a Planet Nine ($a,e$) point as a good average of the behavior that a given Planet Nine would engender.  

\begin{figure*} 
   \centering
   \includegraphics[width=6.5in]{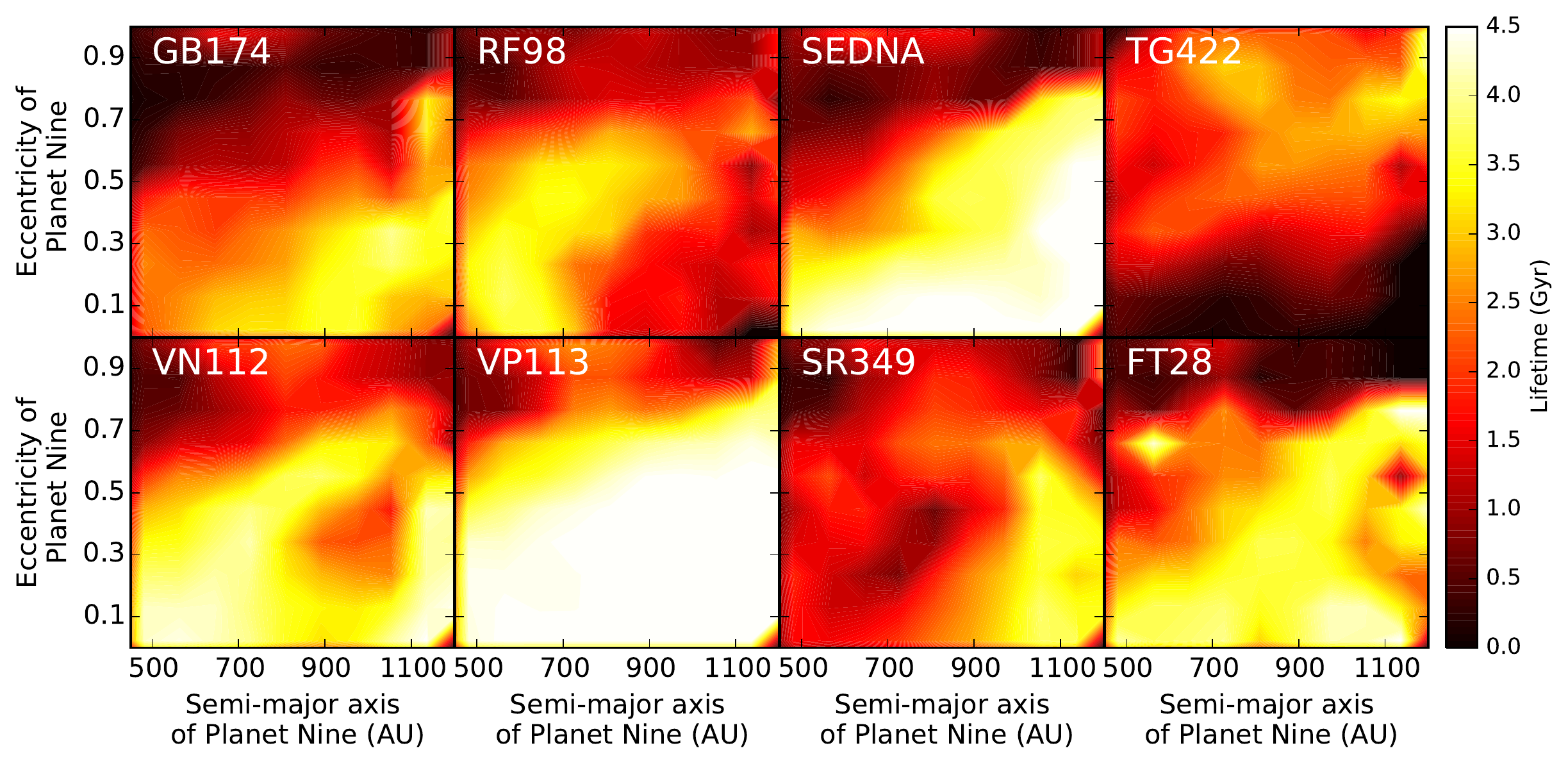} 
   \caption{Individual contour plots of the lifetimes of each of the eight TNOs considered. The lifetime plots do not exhibit the same general trends, indicating that the different objects may be members of different dynamical classes of objects. Table \ref{tab:stab} presents the complete list of their orbital properties; the title of each subplot here is the short name of the object it portrays.}
   \label{fig:stab_map_giant}
\end{figure*}

We construct the contour plots in Figure \ref{fig:stab_map_giant} and Figure \ref{fig:stab_lil} by using a polyharmonic spline \citep[from][]{scipy} to smooth and interpolate between the points generated by our Monte Carlo simulations. We investigated the effects of different interpolation methods, and found that for our sample size ($N\sim1500$ realizations), the interpolation scheme does not drastically affect the final result. However, it is important to note that when we attempted to use $N\sim 100$ realizations, the results varied drastically between different interpolation schemes. Since our sample size does not show such variations between methods, we consider the ($a,e$) parameter space well-sampled.  
Figure \ref{fig:spline} demonstrates the smoothed two-dimensional lifetime function and provides a comparison to the raw points for one object (2012 VP$_{113}$) in our sample. Analogous plots (not presented) can be constructed for each of the eight objects in our sample. 

\subsection{Numerical Results}
\label{sec:numres}
The results of the simulations are plotted in Figure \ref{fig:stab_map_giant}, which presents contour plots for the expected orbital lifetimes of the eight TNOs in the presence of various realizations of Planet Nine. The main results that we get from this experiment fit into three categories: we affirm the potential stability of all eight TNOs in the presence of Neptune and Planet Nine, we find that different TNOs prefer different parameter spaces of Planet Nine, and we examine the fate of the TNOs that do go dynamically unstable.

\textbf{Stability of TNOs.} For all objects in our sample, there are realizations of Planet Nine that allow them to remain dynamically stable for the lifetime of the solar system. Here, dynamical stability requires that a TNO remain within 100 AU of its starting orbit, and not experience collisions with the planets or central body. This value of 100 AU was chosen from examination of the time-evolution of the bulk set of TNOs. We find that when orbits vary by less than 100 AU, the objects are generally confined to a network of mean motion resonances (MMRs), an outcome that we describe in Section \ref{sec:resonance}. 

When orbits change by more than 100 AU, they potentially change dynamical class. The criteria in \citet{bb16} included the cut that $a_{TNO}>250$ AU, and 100 AU of migration inwards would change this dynamical classification of 5 of the 8 TNOs considered in this work. Since this choice of cutoff is somewhat arbitrary, we present in Table \ref{tab:tabtwo} the frequency of final dynamical outcome for each TNO in our sample. Migration is in outcome in between 7\%-30\%, depending on the TNO under consideration, so future work that uses a different criterion for migratory instability can expect variation from our results of this magnitude.

\begin{deluxetable*}{lccccccccc}
\tablecaption{Stability of ETNOs in presence of Neptune and Planet Nine}
\tablewidth{0pt}
\tablehead{
  \colhead{Object Name} &
  \colhead{$a$ (AU)}     &
  \colhead{	$e$	} &
  \colhead{$i$ (degrees)}     &
  \colhead{$\omega$ (degrees)}     &
   \colhead{$\Omega$ (degrees)}     &
  \colhead{Percent Stable}   &
  \colhead{$\bar{\delta t}^{1}$}  &
  \colhead{} 
}
\startdata
2003 VB$_{12}$ (Sedna)	&	499	&	0.85	&	11.92 	& 311.5 & 144.5	&	45.3\%	&	300 Myr &	\\
2004 VN$_{112}$	&	318	&	0.85	&	25.56	& 327.1 & 66.0 	&	28.4\%	&	625 Myr &	\\
2007 TG$_{422}$	&	482	&	0.93	&	18.59	& 285.8 & 113 &	14.6\%	&	1.24 Gyr&	\\
2010 GB$_{174}$	&	371	&	0.87	&	21.54	& 347.8 & 130.6 &	17.9\%	&	1.03 Gyr &	\\
2012 VP$_{113}$	&	261 &	0.69	&	24.06	& 292.8 & 90.8&	56.5\%	&	9 Myr &	\\
2013 RF$_{98}$	&	325	&	0.88	&	29.61	& 316.5 & 67.6 &	18.7\%	&	1.01 Gyr &	\\
2013 FT$_{28}$	&	310	&	0.86	&	17.3	& 40.2 & 217.8 &	19.9\%	&	 709 Myr  &	\\
2014 SR$_{349}$	&	288	&	0.84	&	18.0	& 341.3 & 34.8 &	18.1\%	& 917 Myr	 &	\\
\enddata
\tablecomments{A list of the TNOs used for the dynamical survey. The dynamical stability of each object was evaluated using a suite of numerical N-body simulations. The orbital elements of each object provided in this table are the best fit observational values. In the simulations, the orbital elements were drawn from the 1$\sigma$ distribution for each realization of each object. Also provided in this table is the percentage of realizations of each object that are dynamically stable. It is important to note that this percentage reported (which is the percentage of realizations of each object that are dynamically stable) is marginalized over all realizations of Planet Nine included in the simulations, and thus the exact percentage is not meaningful. What is meaningful is that the percentages are all non-zero, indicating that for a selection of orbital parameters and Planet Nine realizations, each object in our sample can be dynamically stable. $^{1}$: $\bar{\delta t}$ is the difference in median lifetime (over all Planet Nine realizations) between two cases: case one being when the definition of dynamical instability does not include migration, and case two being when migration of more than 100 AU constitutes dynamical instability. Larger values indicate that TNOs are susceptible to significant ($\delta a >100$) migrations in semi-major axis.    }  
\label{tab:stab}
\end{deluxetable*}

Table \ref{tab:stab} provides a `Percent Stable' column, which gives for each object the percentage of our 1500 realizations (which all have different $a,e$ for Planet Nine) which allowed that object to remain dynamically stable for the entire 4.5 Gyr simulation. 
Since our simulations also include Neptune as an active particle, all of these objects \emph{can} be dynamically stable in the presence of both Neptune and Planet Nine. Notably, 18 of our 1500 realizations of Planet Nine allowed all eight TNOs to remain dynamically stable for 4.5 Gyr. 
The 18 trials that allowed all tested TNOs to survive had semi-major axes varying between 600 AU and 800 AU and eccentricities between 0.35 and 0.55. The small sample ($N\sim18$) that allowed all TNOs to survive limits any further conclusions based on their orbital parameters. 
\citet{ST_tnos} found that 2007 TG$_{422}$ and 2013 RF$_{98}$ were both dynamically unstable in the presence of Neptune, but they did not include a potential Planet Nine in their integrations. We find that 2007 TG$_{422}$ and 2013 RF$_{98}$ have stability percentages of less than 20\%, making them less stable on average than, for example, 2012 VP$_{113}$. However, with a dynamically favorable realization of Planet Nine, both of these objects can remain dynamically stable for a solar system lifetime. This result, combined with the result from \citet{ST_tnos} that these two objects are not dynamically stable in the presence of Neptune alone, suggests that Planet Nine can boost the orbital stability of these objects. 

\textbf{Variations between behavior of different TNOs.} For each object in our sample, different realizations of Planet Nine lead to differing object lifetimes. This is shown in Figure \ref{fig:stab_map_giant}, which plots the lifetime of each TNO as a function of the semi-major axis and eccentricity of Planet Nine. These stability maps look very different for different objects. For example, large-$a$, low-$e$ orbits lead to a longer lifetime for 2012 VP$_{113}$. For this object, shorter object lifetimes occur if Planet Nine has a shorter perihelion distance (lower $a$ or higher $e$). This is intuitive. In contrast, 2007 TG$_{422}$ is dynamically unstable in the presence of these same long perihelion-distance objects that 2012 VP$_{113}$ preferred. Since different objects prefer different regions of Planet Nine's possible parameter space, a better understanding of the constraints they provide can be obtained by considering all objects simultaneously. We will discuss how this can be done, as well as find constraints using our results thus far, in Section \ref{sec:bayes}.

\textbf{Fate of the TNOs.} To construct the lifetime contours in Figure \ref{fig:stab_map_giant}, we defined the lifetime of an object as the length of time it lived in our simulation without significant alterations in its orbit. We defined significant alterations in orbit to be any of the following: (1) migration in semi-major axis by more than 100 AU from the starting orbit of the object, (2) collisions or a close encounter (defined as passing within 3 Hill radii of the larger body) with Planet Nine or Neptune, (3) collision with the central body, and (4) ejection from the solar system. It is clear that each of these criteria have different thresholds for importance, as well as different outcomes for the object. In particular, while ending methods (2-4) result in a violent end for the TNO, method (1) does not necessarily remove the TNO from the solar system: instead, the TNO's orbit is significantly altered, but the TNO may continue to be a part of the solar system, merely in a different location. Numerical simulations allow these effects to be disentangled. In particular, we find that the two objects (2007 TG$_{422}$ and 2013 RF$_{98}$) that do not prefer the high-$a$, low-$e$ realizations of Planet Nine tend to exhibit the migratory end in those realizations, rather than experiencing a violent instability. We will further discuss this effect in Section \ref{sec:neptune}.

\begin{deluxetable*}{lccccccccc}
\tablecaption{Outcomes of ETNOs in presence of Neptune and Planet Nine}
\tablewidth{0pt}
\tablehead{
  \colhead{Object} &
  \colhead{Migration ($a>100$ AU)} &
  \colhead{Close Encounter}     &
  \colhead{	Collision with Central Body} &
  \colhead{Ejection}     &
  \colhead{Survive to 4.5 Gyr}     &
}
\startdata
2003 VB$_{12}$ (Sedna)	&	16.0\%	&	19.3\%	&	0.1\%	&	19.3\%	&	45.3\% \\
2004 VN$_{112}$	&	14.9\%	&	30.6\%	&	0.7\%	&	25.4\%	&	28.4\% \\
2007 TG$_{422}$	&	29.3\%	&	32.9\%	&	0.0\%	&	23.1\%	&	14.6\% \\
2010 GB$_{174}$	&	17.8\%	&	31.2\%	&	0.4\%	&	32.7\%	&	17.9\% \\
2012 VP$_{113}$	&	7.1\%	&	21.1\%	&	0.5\%	&	14.8\%	&	56.5\% \\
2013 RF$_{98}$	&	14.0\%	&	40.6\%	&	0.0\%	&	26.7\%	&	18.7\% \\
2013 FT$_{28}$	&	17.2\%	&	30.7\%	&	0.0\%	&	32.2\%	&	19.9\% \\
2014 SR$_{349}$	&	22.0\%	&	39.5\%	&	0.7\%	&	19.8\%	&	18.1\% \\
\enddata
\tablecomments{For each TNO used in the dynamical simulations, the percentage of integrations that ended in each major instability outcome.   }  
\label{tab:tabtwo}
\end{deluxetable*}

\section{Deriving constraints on the orbital elements of Planet Nine}
\label{sec:bayes}
Given the predicted lifetime distributions constructed in the previous section, we can constrain the orbital properties of Planet Nine by determining which ($a,e$) combinations allow the continued dynamical stability of the observed clustered TNOs. We make an assumption here that in order for the TNOs to attain their observed $\varpi$ clustering, they must have remained in their currently observed orbits for at least secular timescales. This would require those orbits being dynamically stable in the presence of Planet Nine. The process of constructing these constraints will be detailed in this section. 

\subsection{Bayesian inference towards a posterior probability distribution for the orbital elements of Planet Nine}
The ultimate goal of our dynamical stability analysis is to determine the posterior probability distribution for the orbital elements ($a,e$) of Planet Nine. This will serve as a check and supplement to previous orbital posteriors in the literature, which were estimated using different techniques. We define $A$ to be the orbital elements of Planet Nine, $T_{i}$ to be the expected dynamical lifetime of the $i^{th}$ TNO, and $I$ to be all prior information that we can incorporate into our models (for example: the fact that Planet Nine exists is a prior we impose, as are the observational errors of the discovered TNOs, which lead to uncertainties on their orbits). With these definitions, the property we wish to measure is $P(A|T_{i}, I)$, the probability of $A$ (Planet Nine's orbital elements), given the lifetimes of the observed TNOs and our other knowledge. 
This posterior probability function can be represented using Bayes' Theorem as follows:
\be
P(A|T_{i}, I) = \dfrac{P(T_{i}, I|A)P(A)}{P(T_{i}, I)}  
\ee
However, upon inspection of this expression, there is one clear problem: $P(T_{i}, I|A)$ requires knowledge of $T_{i}$, the actual lifetimes of the observed TNOs in our solar system. This is not a property that can be measured. As an added difficulty, the prior information encapsulated in $I$ includes large uncertainties on the orbital elements of the observed TNOs. 

To overcome this difficulty, we define a new parameter, $D_{i}$. $D_{i}$ is defined to be the computed lifetimes of the \emph{observed} TNOs, where these lifetimes are marginalized over the observational priors of the TNOs' orbital elements and our other uncertainties that were before folded into $I$. Although we cannot measure the true lifetimes $T_{i}$, we have constructed TNO lifetime estimates $D_{i}$ as they depend on our priors by using numerical N-body simulations. These simulations are described in the previous section, and a visualization of the derived lifetimes is presented in Figure \ref{fig:stab_map_giant}.

Now, we can rewrite the Bayesian statement of our posterior in terms of this new variable:
\be
P(A|D_{i}) = \dfrac{P(D_{i}|A)P(A)}{P(D_{i})} 
\ee
when $P(D_{i}|A)$ is the probability of getting the numerically measured lifetimes conditional on the orbital elements of Planet Nine, $P(A)$ is our priors on Planet Nine's orbital elements, and 
$P(D_{i})$ is the occurrence probability of our computed lifetimes. 
Of course, we have $N$ TNOs to consider in this analysis, so what we really need to compute is the posterior probability distribution as it depends on the numerically estimated lifetimes of all $N$ objects: 
\begin{equation*}
P(A|D_{1}, D_{2}... D_{N}) = \dfrac{P(D_{1}, D_{2}... D_{N}|A)P(A)}{P(D_{1}, D_{2}... D_{N})},
\end{equation*}
which, using the definition of Bayes' theorem, reduces into the form:
\begin{equation}
\begin{split}
P(A|D_{1}, D_{2}... D_{N}) = \dfrac{P(D_{1}) \times P(D_{2})\times ... P(D_{N})}{P(D_{1}, D_{2}... D_{N})}\times \\
\dfrac{P(D_{1}|A) \times P(D_{2}|A)\times ... P(D_{N}|A)}{P(A)^{N-1}}
\end{split}
\label{eq:12}
\end{equation}
We can say that $D_{1}, D_{2}...D_{N}$ are conditionally independent, since we treated all TNOs as test particles in the simulations used to generate $\{D_{i}\}$. Since each TNO has zero mass in the simulations, removing one will not alter the lifetime maps for the other objects. As a result, the first term of the right-hand side of Equation \ref{eq:12} can be treated as a normalization coefficient, which is needed only when comparing different versions of $P(A|D_{1}, D_{2}... D_{N})$ constructed with different numbers $N$ of TNOs. We need only a relative posterior probability distribution, which will identify the most likely realization of Planet Nine and identify the parameter space in which it is likely to reside.  

The final result we need to compute, then, is:
\begin{equation}
P(A|D_{1}, D_{2}... D_{N}) = a \dfrac{\prod_{n=1}^{n=N} P(D_{i}|A)}{P(A)^{N-1}}
\label{eq:final}
\end{equation}
when $D$ denotes the lifetime of a TNO in the presence of Planet Nine as measured from simulations, $i$ denotes the TNO considered, $A$ denotes the orbital elements ($a,e$) of Planet Nine. $P(A)$ is a known quantity, and $P(D_{i}|A)$ can be measured from the stability maps constructed using the N-body integrations.

\subsection{Converting TNO lifetime maps to probability distributions}
Equation \ref{eq:final}, which can be used to construct our goal posterior probability distribution, requires $P(D_{i}|A)$ for each TNO in the sample. $P(D_{i}|A)$ is the probability of the TNOs' lifetimes as they depend on the orbital elements of Planet Nine. To compute this term, we must convert the lifetime maps presented in Figure \ref{fig:stab_map_giant} into probability distributions.

\begin{figure} 
   \centering
   \includegraphics[width=3.4in]{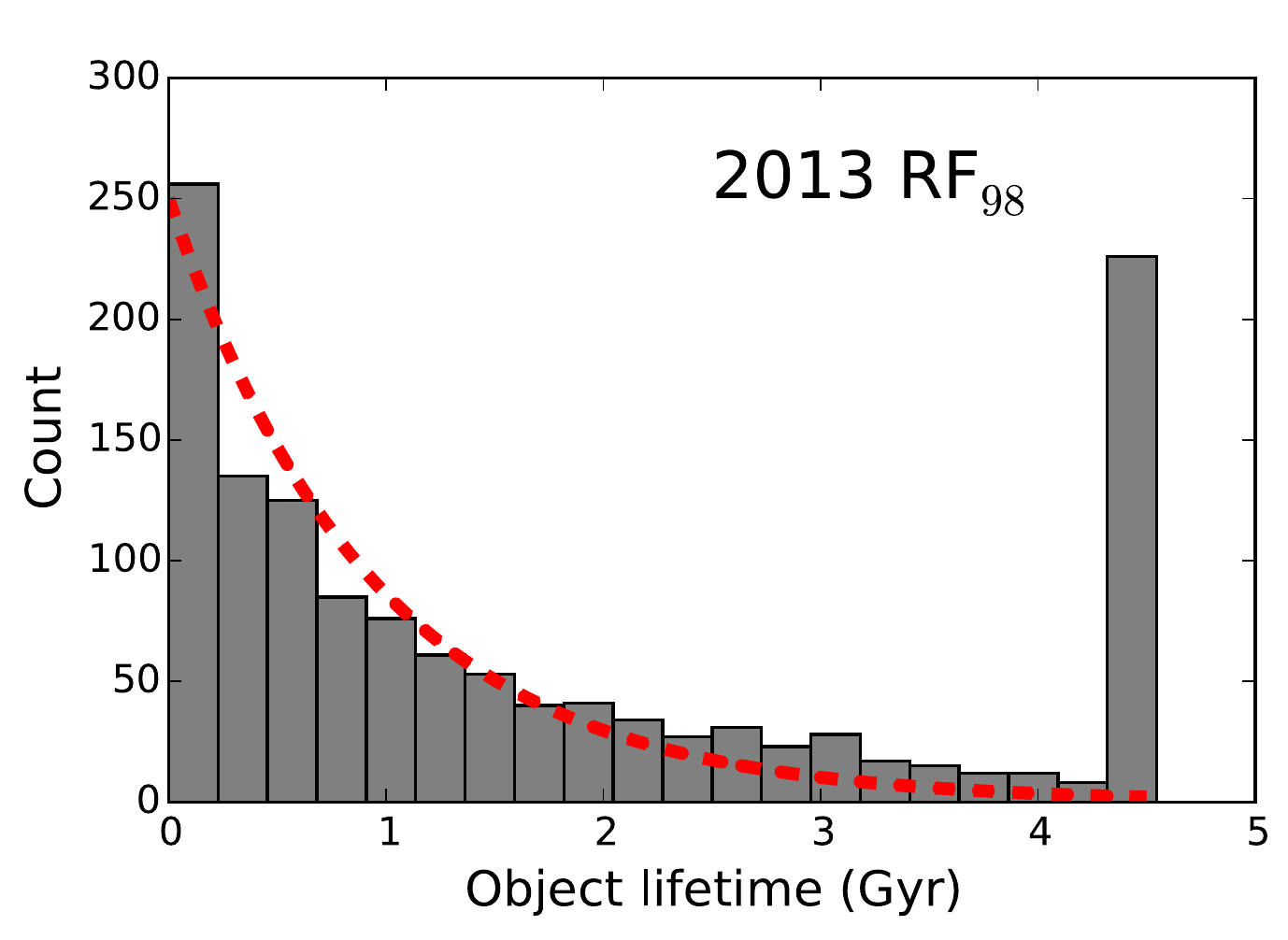} 
   \caption{ A histogram of lifetimes (measured from time $t=0$ at the start of the N-body simulation) for 2013 RF$_{98}$. Barring a pile-up at lifetimes of $t=4.5$ Gyr (corresponding to realizations that were stable for the duration of the simulation; these were not included in the fit), the decay appears to have an approximately exponential trend.   }
   \label{fig:histo}
\end{figure}

Each realization of Planet Nine will lead to a different time of dynamical instability (defined as the time at which an object experiences one of the four instability mechanisms described in Section \ref{sec:numres}), with some objects never experiencing dynamical instability. 
In Figure \ref{fig:histo}, we plot a histogram of the object lifetimes for 2013 RF$_{98}$, as derived from the $1500$ numerical integrations run with varying realizations of Planet Nine.

Although the longer object lifetimes can intuitively be interpreted as corresponding to more likely realizations of Planet Nine, we would like to convert these lifetimes into a relative probability function. 
Since an exponential decay trend appears to be a good, empirical fit for the lifetimes of an object marginalized over all integrations (regardless of the fact that the realization of Planet Nine varies between the different trials), we can fit a decay constant $\lambda$ from the lifetime histogram for all decaying realizations of a single TNO:
\be
f_{\rm{decay}}(t) = C_{0} e^{-\lambda t}
\ee
when $f_{\rm{decay}}(t)$ is the fraction of realizations decaying at each time $t$, $C_{0}$ a normalization constant, and $\lambda$ the decay constant to be fit. 
For each object in our sample, the histogram of instability times was fit with exponential curve using a simple Levenburg-Marquardt optimization scheme, with all objects that do not experience dynamical instability (which have lifetimes $t=4.5$ Gyr) excluded from the fit. The results of this fit for 2013 RF$_{98}$ are plotted in Figure \ref{fig:histo}. For this particular object, if we extend the exponential curve to infinite time, we expect only 0.4\% of these dynamically stable objects to become dynamically unstable, indicating that the majority of objects that have not decayed after 4.5 Gyr are truly dynamically stable. 
 
To construct a final probability distribution, we need to account for the fact that for each object, some fraction of trials were dynamically stable for the entire 4.5 Gyr integration length. For this reason, we construct a piecewise probability function for $P(D_{i}|A)$, the probability we get these lifetimes from our simulation given a particular set of $A$, Planet Nine's orbital elements:
\begin{equation}
P(D_{i}|A) =  \begin{cases}
N_{s}/N   &\text{if D$_{i}$=4.5 Gyr}\\
C_{1} \times e^{-\lambda D_{i}} &\text{else}
\end{cases}
\label{eq:piecewise}
\end{equation}

when $P(D_{i}|A)$ is the probability of computing the object lifetime $D_{i}$ for the i$th$ TNO in the presence of a Planet Nine with orbital elements $A$, $N_{s}$ the number of integrations for which the object did not experience dynamical instability, $N$ the total number of integrations, $\lambda$ the decay constant determined from the previous fit, and $C_{1}$ is a normalization constant of the form:
\be
C_{1} =  \frac{N - N_{s}}{N}\times \left[ \int^{4.5\ \rm{Gyr}}_{0} e^{-\lambda t} dt \right]^{-1}
\ee
The substitution of Equation \ref{eq:piecewise} into Equation \ref{eq:final} results in a final expression of the posterior probability distribution $P(A|D_{1}, D_{2}... D_{N})$. For the eight objects considered in this work, this final distribution is plotted in Figure \ref{fig:stab_lil}. 
It is important to note that this process required multiple steps of normalization, which depend on $P(A)$, the priors, and the populations of TNOs used to derive constraints. Since we (by necessity) treated the objects used in this analysis as conditionally independent, the posterior presented in Figure \ref{fig:stab_lil} provides relative, rather than absolute, measures of the likelihood of each ($a,e$) realization of Planet Nine. 
The derived decay constant must be re-derived if different populations of TNOs or different orbits of Planet Nine are tested in the future; the values used in this work are particular to our sample of eight TNOs in the presence of our particular population of Planet Nines. For this reason, we do not provide the decay constants in this work, as they cannot be used for these objects in general but only for our particular choices of Planet Nine's priors.

\subsection{The posterior probability distribution for Planet Nine's orbital elements}
\begin{figure} 
   \centering
   \includegraphics[width=3.4in]{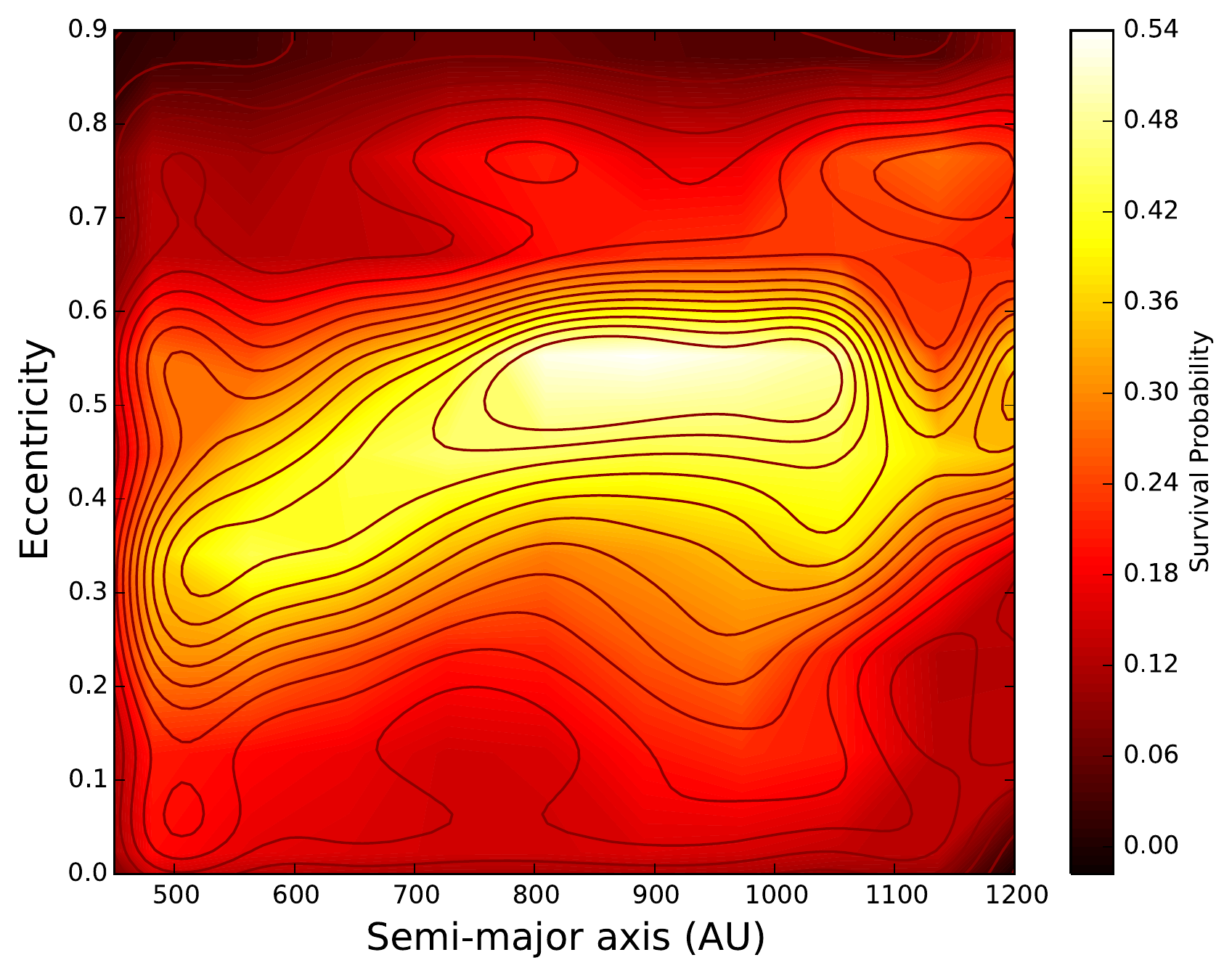} 
   \caption{The overall stability posterior for the semi-major axis and eccentricity of Planet Nine. This posterior was constructed by taking a summation of the posteriors for each individual object, including the six objects used in \cite{bb16} and the two new high-$a$, low-$e$ objects from \citet{ST_tnos}. }
   \label{fig:stab_lil}
\end{figure}
Figure \ref{fig:stab_lil} presents the final posterior probability distribution for Planet Nine's orbital elements ($a,e$), based on the observed dynamical stability of the eight TNOs considered in this work. This distribution provides a relative measure of the likelihood of differing combinations of Planet Nine's semi-major axis and eccentricity. 
The nominal orbit of Planet Nine (700 AU, 0.6 eccentricity) appears to lie in a less preferred region, with slightly smaller eccentricities (0.3-0.4) being preferred. 
Remarkably, the large-$a$, low-$e$ orbits appear to be excluded based on our dynamical stability arguments. This is roughly consistent with the posterior generated using clustering arguments, provided in \citet{bb162}, which also preferred a middling eccentricity and excluded large-$a$, low-$e$ iterations of Planet Nine. 

\subsection{Comparing our prediction of Planet Nine's Orbit with constraints derived via different methods}
\label{sec:compare}
\begin{figure*} 
   \centering
   \includegraphics[width=6.8in]{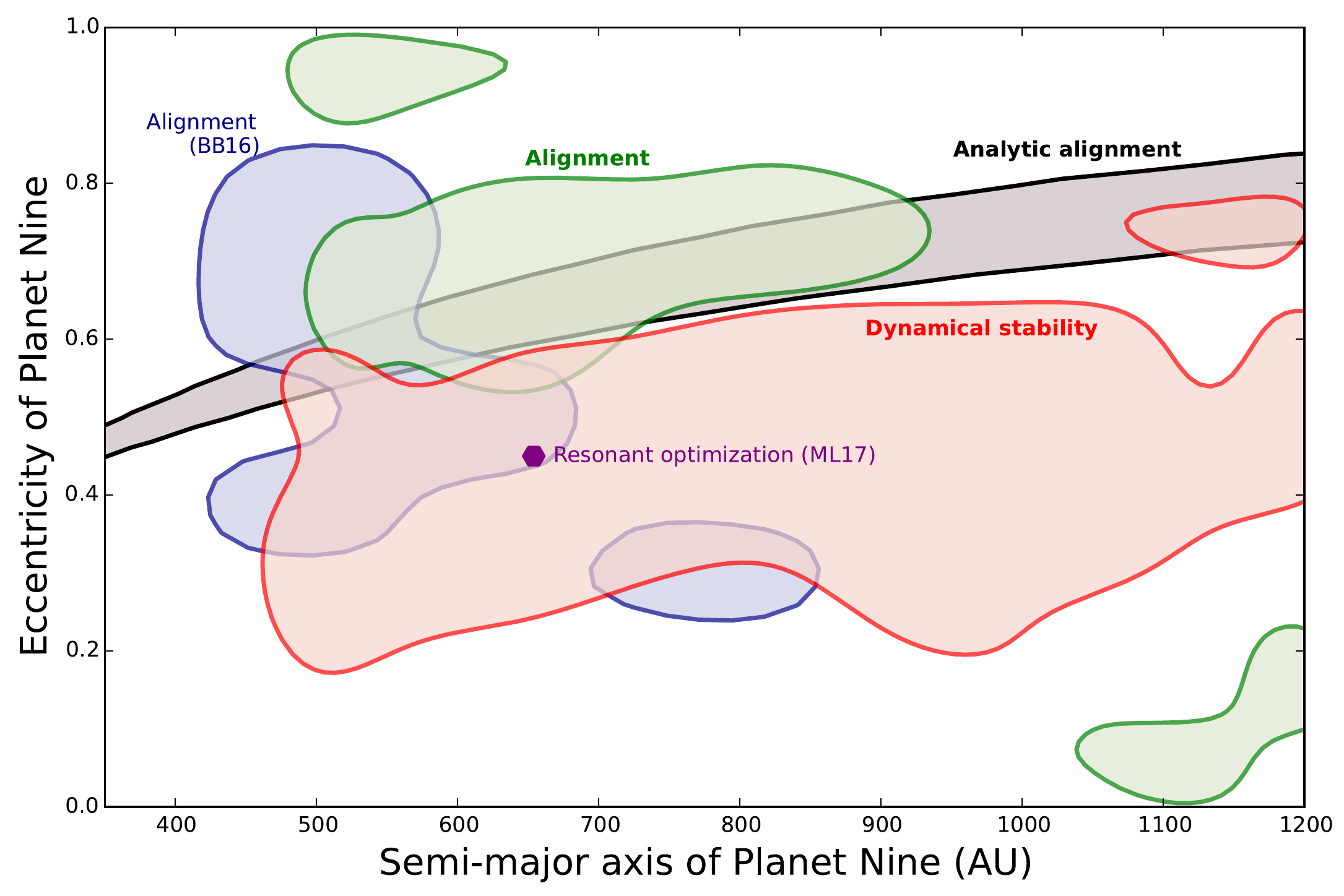} 
   \caption{A comparison between the preferred regions for Planet Nine's orbit, as computed using a variety of different methods. Bold labels indicate that the posterior was derived in this work. In red, we plot the region that maximizes the survival probability for the TNOs in our N-body simulations (based on their dynamical stability; see Figure \ref{fig:stab_lil}). In green, we plot the realizations of Planet Nine orbital parameters in our N-body simulations that allowed the TNOs to be aligned as observed in nature. In black, we plot the analytic approximation for the region where we expect the observed alignment to occur (see Equation \ref{dwdt} and its derivation in the Appendix \ref{app2}). In blue, we plot the region reported in \citet{bb162} that allows the alignment of TNOs in a suite of numerical N-body simulations (unlike the green curve, these simulations used randomized test particles). The purple point denotes the best-fit orbit as reported by \citet{sarah}, which was found by optimizing the resonant behavior of the TNOs in the presence of Planet Nine. Each method prefers somewhat different regimes of parameter space, with the dynamical stability argument allowing the largest region. There is significant overlap between the results derived from the different methods. The optimal orbital elements for Planet Nine exist in the overlap region, which corresponds to eccentricity in the range $e_9$ = 0.4 -- 0.6 and semimajor axis in the range $a_9$ = 500 --700 AU.  }
   \label{fig:test}
\end{figure*}

The posterior probability distribution we present in Figure \ref{fig:stab_lil} was generated using dynamical stability arguments. Specifically, Planet Nine realizations that allow the observed TNOs to remain dynamically stable are considered to be more likely than those that cause dynamical instabilities. \citet{bb162} also provided a probability map showing the most likely regions for Planet Nine based on the observed orbital alignment of the TNOs. In Figure \ref{fig:test}, we compare the two probability distributions presented in \citet{bb162} to the distribution derived in this work (along with additional contours described below). For these distributions, we plot a single contour in Figure \ref{fig:test} using the value $1\sigma$ below the maximum probability of the distribution. The level of this contour is chosen so that the single contours visually reflect the highest probability regions for each posterior type. The choice of the (maximum -  $1\sigma$) level is representative, but by no means the only way to visualize the comparisons between the posterior types. 

The regions showing the most overlap between the 10 $M_{\oplus}$ alignment result from \citet{bb162} and our dynamical stability result are those with a semi-major axis between 525--675 AU and an eccentricity of 0.40--0.55, and the region with a semi-major axis between 700--800 AU and an eccentricity of 0.25--0.35. A recent paper by \citet{sarah} also used dynamical stability arguments and resonance considerations to choose a best-fit ($a,e$) of (654 AU, 0.45). This point is also plotted in Figure \ref{fig:test} and appears to be consistent with the region showing overlap between both the dynamical stability results of this work and those of \citep{bb162}. 

The posteriors plotted in Figure \ref{fig:test} include four curves based on N-body simulations and one analytical curve. The analytical curve encompasses the regime where the equation of motion $d\omega / d t \approx$ 0. It is important to note that this analytic approximation is not a true analysis of the dynamics of the system, as the derivation (the complete form of which is presented in Appendix \ref{app1}) assumes all bodies are coplanar, and solves for alignment rather than anti-alignment. Thus, this analytic model is more an order-of-magnitude estimate of the effect we expect to see, rather than a true prediction. 

The functional form for $d\omega / d t$ is derived from the secular Hamiltonian, and assumes the form:
\be
\begin{aligned}
\frac{d\omega}{dt} & \propto  \frac{3}{4 M_{c}} \sum^{4}_{i=1} m_{i}a_{i}^{2} (e^{2} -1)^{-2} a^{-2} \\
- & \frac{15}{16} \frac{m_{9}}{M_{c}} \frac{e_{9}}{e} 
\left( \frac{a}{a_{9}} \right)^{4} \frac{\sqrt{1-e^{2}}}{(1-e_{9}^{2})^{5/2}} 
(1- 9e^{2}/4) \cos(\omega - \omega_{9})\\
- & \frac{3}{4} \frac{m_{9}}{M_{c}} 
\left( \frac{a}{a_{9}} \right)^{3} \frac{\sqrt{1-e^{2}}}{(1-e_{9}^{2})^{3/2}},
\label{dwdt}
\end{aligned}
\ee
where the subscript $9$ denotes the orbital elements of Planet Nine, the subscript $i$ within the summation in the first term denotes the four giant planets, $c$ denotes the central body, and a lack of subscript denotes the TNO for which the equation of motion is written. This equation of motion can be split into two parts: $d\omega_{SS} / d t$, which denotes precession due to the effect of the gas giants, and  $d\omega_{9} / d t $, which denotes the precession due to the effect of Planet Nine. The magnitudes of these terms can be written in the form 
\begin{equation}
\begin{aligned}
\frac{d\omega_{SS}}{dt} \propto& \frac{3}{4 M_{c}} \sum^{4}_{i=1} m_{i}a_{i}^{2} (e^{2} -1)^{-2} a^{-2} \\
\frac{d\omega_{9}}{dt} \propto&  \frac{15 e_{9} }{16 e }\frac{m_{9}}{M_{c}} \left( \frac{a}{a_{9}} \right)^{4} \frac{\sqrt{1-e^{2}}}{(1-e_{9}^{2})^{5/2}} 
(1- \frac{9e^{2}}{4}) \cos(\omega - \omega_{9}) \\
& + \frac{3}{4} \frac{m_{9}}{M_{c}}
\left( \frac{a}{a_{9}} \right)^{3} \frac{\sqrt{1-e^{2}}}{(1-e_{9}^{2})^{3/2}}
\end{aligned}
\label{eq:two}
\end{equation}
where $d\omega_{TNO} / d t = d\omega_{SS} / d t - d \omega_{9} / d t$. At the point where $d\omega_{SS} / d t \approx d \omega_{9} / d t$, the precession rates due to Planet Nine and the inner solar system cancel each other out, and the TNO orbit is not expected to precess. For each TNO, we can construct a curve in the ($a,e$) plane for which these precession rates cancel. The region presented in Figure \ref{fig:test} is the superposition of these curves for all eight TNOs, and it encompasses the range of Planet Nine realizations which will allow alignment by preventing precession. 

Figure \ref{fig:test} also presents (in green) a posterior based on alignment of the TNOs, as derived from our set of numerical simulations. Our simulations were intended to test dynamical stability, but one output of the N-body simulation is the orbital elements $\omega$ and $\Omega$ of the TNOs over time. Using a similar technique to that used to construct the dynamical stability posterior in Figure \ref{fig:stab_lil}, we tested how well aligned the TNOs were for each realization of Planet Nine. Alignment was measured by looking at the fraction of time for which all eight TNOs were aligned with each other (and thus anti-aligned with Planet Nine: we counted dispersions of less than 94 degrees in $\varpi = \omega +\Omega$ between all eight TNOs as aligned). We constructed a contour plot of the percentage of the integrations during which alignment was visible for all eight TNOs, which exhibited between near 0\% alignment to a maximum of $\sim$10\% alignment in the green regions. The alignment rate expected for pure chance would be about $2\times10^{-5}$ (for alignment among all eight TNOs). As a result, the numerical simulations show an overabundance of alignment in certain, preferred regions (denoted by the green contour of Figure \ref{fig:test}).

The constraints provided by the three methods in this work and the two literature results show rough agreement: the region of Planet Nine parameter space looking most attractive extends
over the range 500--700 AU in semi-major axis, and 0.3-0.6 in eccentricity. Notably, low eccentricities are disfavored. Notice also that the contours preferred for TNO stability are somewhat parallel to those for orbital alignment. In order for Planet Nine to be close enough to the TNOs to enforce alignment, it must be close enough that the TNO orbit is close to instability. As a result, the preferred region is given by the boundary between the stablity contours and the alignment contours in Figure \ref{fig:test}. 

\section{End States for Trans-Neptunian objects under the influence of Planet Nine}
\label{sec:neptune}
The numerical simulations that we use to create distributions of TNO lifetime in the presence of Planet Nine evaluate the long-term evolution of these TNOs. As discussed in Section \ref{sec:num_methods}, there are five main outcomes for these TNOs: (1) migration in semi-major axis, which is defined in the simulations as the orbit's semi-major axis attaining a value more than 100 AU from the starting orbit; (2) a close encounter (defined as passing within 3 Hill radii of the larger body) with Planet Nine or Neptune, which would result in the TNO being captured by or colliding with the large planet; (3) collision with the central body; (4) ejection from the solar system, where the ejection radius is taken to be 10,000 AU\footnote{This choice in ejection radius will remove objects which become unbound; it is important to note that there may be second order effects due to stellar encounters \citep[see, for example,][]{star9}. The effect on TNO motion due to these external perturbations is expected to be small, and is neglected in this work.}; and finally, (5) dynamical stability, where none of the aforementioned effects occur over the 4.5 Gyr timescale of the simulation. This final result (5) corresponds to scenarios that are consistent with observations, where the TNO can remain in its orbit over solar system lifetimes. This scenario allows the secular alignment of the TNOs' longitude of perihelion by Planet Nine. However, if this last result does not occur, then the TNO will experience one of the first four (1-4) outcomes. 

Within these four dynamical instability outcomes, there is further stratification in the effects of each mechanism. The violent ends (collision with a planet, the star, or ejection from the system) generally remove the TNO from the solar system entirely, while the migratory outcome is more nuanced. 
TNOs that are in the process of migrating cannot necessarily be used as evidence of the Planet Nine hypothesis, since their orbital elements are in flux and may not have had sufficient time in their current orbits to attain the alignment that serves as a hallmark of Planet Nine's influence.

2007 TG$_{422}$ was one of the more interesting objects in the sample of TNOs considered in this work. Although found in \citet{ST_tnos} to be dynamically unstable in the presence of Neptune, we reevaluated the stability of 2007 TG$_{422}$ in the presence of Planet Nine in Section \ref{sec:num_methods}, and found that adding Planet Nine to the solar system can actually stabilize the orbit of this object. 2007 TG$_{422}$ has a particularly high semi-major axis and eccentricity, suggesting that its dynamical instability might be due to orbit crossing with other planets. However, this is not the case. The most common dynamical instability mechanism for 2007 TG$_{422}$ is actually migration, leading this object to remain present in the solar system but wander from its starting orbit. This migratory outcome is likely what \citet{ST_tnos} found in their work, and we reproduce this instability in the case of large-$a$, low-$e$ realizations of Planet Nine (which is dynamically very similar to the case of no Planet Nine, as considered in \citet{ST_tnos}). 

\begin{figure} 
   \centering
   \includegraphics[width=3.4in]{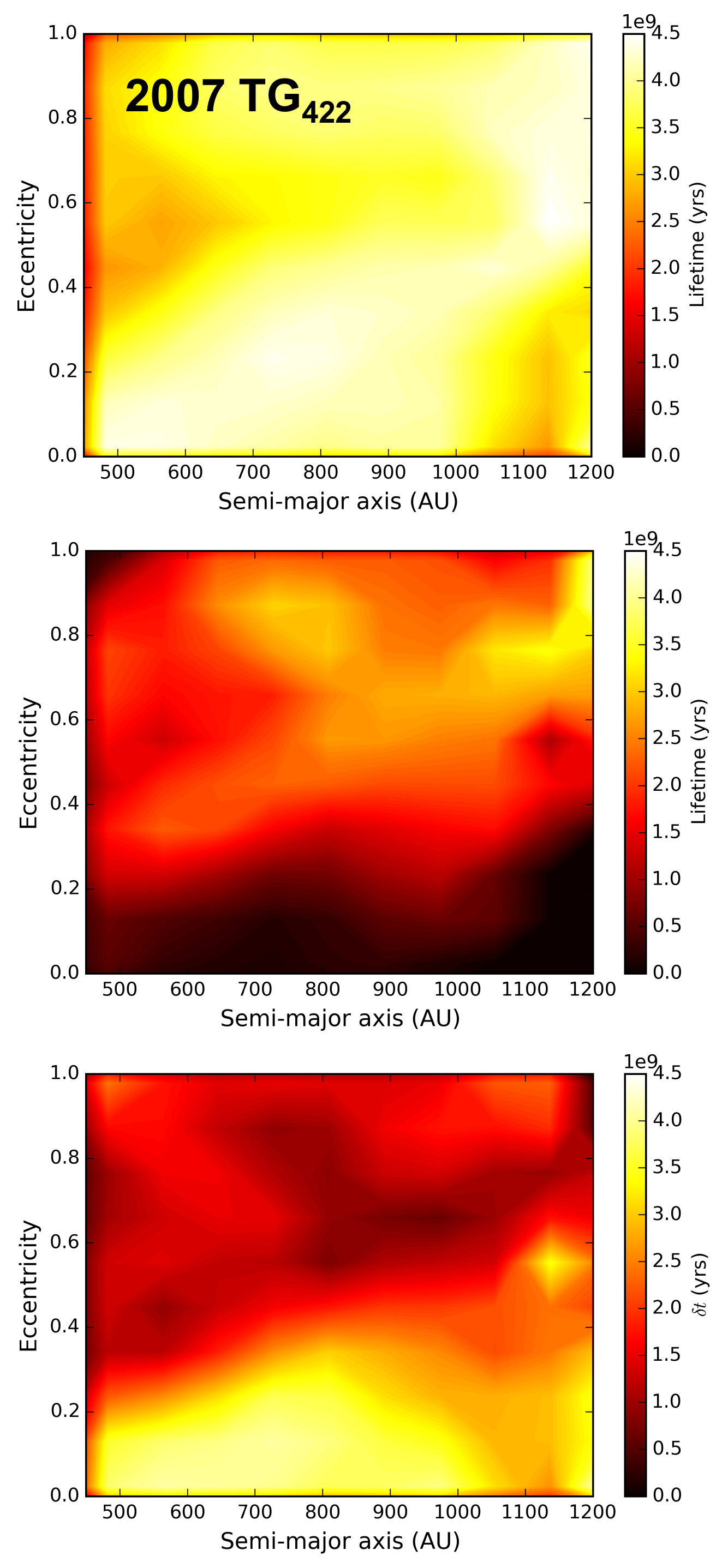} 
   \caption{ The lifetime of 2007 TG$_{422}$ in the presence of realizations of Planet Nine with semi major axis between 450 and 1200 AU, and eccentricities between 0 and 1. \emph{(top panel)} Stability lifetimes when dynamical instability is only caused by violent ends for 2007 TG$_{422}$, including collision with a solar system planet, collision with the sun, or ejection from the solar system (the outer boundary of which is defined to be 10000 AU). \emph{(middle panel)} Stability lifetimes, when dynamical instability is caused by the violent end depicted in the top panel and also migration in semi-major axis by more than 100 AU. \emph{(bottom panel)} The difference $\delta t$ in stability lifetimes between the two cases, demonstrating which realizations of Planet Nine are most likely to cause 2007 TG$_{422}$ to change its orbit significantly. Planet Nine realizations with low eccentricity and large semi-major axis cause 2007 TG$_{422}$'s orbit to migrate, but not meet with a violent end.   }
   \label{fig:tg422}
\end{figure}

The top panels of Figure \ref{fig:tg422} show the instability lifetime map for 2007 TG$_{422}$ when only violent instability methods are considered (top panel) and when migratory instabilities are also considered (middle panel). The significantly shortened lifetimes present in this middle panel show that migration is the main explanation for 2007 TG$_{422}$'s dynamical instability in the presence of Planet Nine. The bottom panel of Figure \ref{fig:tg422} shows the difference in expected dynamical stability lifetime for this object between the cases depicted in the top two panels: when this difference is low, migration does not have a large effect in ending the integrations, and when it is high, migration is a very common outcome for 2007 TG$_{422}$ in the presence of that particular Planet Nine realization. This difference plot highlights the regions where migration acts as the main cause of dynamical instability. The large-$a$, low-$e$ realizations of Planet Nine tend to cause migration of 2007 TG$_{422}$, but not violent dynamical instabilities.  

The susceptibility of an object to migration in the presence of Planet Nine can also be summarized as $\bar{\delta t}$, the difference in median lifetime (over all Planet Nine realizations) for an object between the case when migration is considered to be a dynamically unstable outcome and when it is ignored. Larger values of $\bar{\delta t}$ indicate that the TNO is more susceptible to significant migration in semi-major axis ($\delta a > 100$ AU), and that such migrations significantly change the dynamical stability map of that object. The choice of $\delta a >100$ AU as the threshold criterion for migration is somewhat arbitrary. It is chosen to represent the condition for which an object has drifted significantly from its observed orbit. Migration in semi-major axis with deviations greater than 100 AU generally results in an orbit with significantly different orbital elements, rather than oscillation of the orbital elements around well-defined central values. As a result, objects that migrate by more than 100 AU are unlikely to remain part of the same dynamical class of object.

For 2007 TG$_{422}$ , $\bar{\delta t} = 1.07$ Gyr, meaning that Planet Nine incites migration in this object often enough to decrease its dynamically stable lifetime by more than one billion years. In contrast, 2012 VP$_{112}$ has a $\bar{\delta t} = 32$ Myr, indicating that migration does not play a large role in its outcomes. 
The values of $\bar{\delta t}$ for each object in our sample are reported in Table \ref{tab:stab}, which presents in its last two columns both $\bar{\delta t}$ and the percentage of trials out of all our simulations that are dynamically stable. As the percentage increases (or, as an object is more stable in the presence of any considered Planet Nine realization), the value of $\bar{\delta t}$ tends to decrease (or, the TNO experiences migration as an instability outcome with a lower frequency). This suggests that susceptibility to migration is a major factor leading to differences in orbit lifetimes between different TNOs in the presence of Planet Nine. 

In this work, we have made the assumption that since the eight TNOs in our sample exhibit alignment in longitude of perihelion attributable to Planet Nine, these objects by necessity have lived in their current orbits for a significant length of time. In constructing the stability posterior presented in Figure \ref{fig:stab_lil}, we assumed that since significant, unbounded migration alters orbits and potentially disrupts this alignment, a migratory instability provides equal information against Planet Nine's orbital elements as does a violent dynamical instability. In this way, we used migratory instabilities (such as those caused by large-$a$, low-$e$ realizations of Planet Nine for 2007 TG$_{422}$) as evidence against the iterations of Planet Nine that excited the migration. 

The TNOs affected by Planet Nine appear to fall into two categories: some of them (Sedna, 2012 VP$_{113}$) are generally stable against this migratory process, while others (2007 TG$_{422}$, 2013 RF$_{98}$) can be easily caused to migrate in semi-major axis with an improperly chosen realization of Planet Nine. The best orbit of Planet Nine can be determined not only from the orbit-crossing constraint that rules out small-$a$, large-$e$ orbits for all TNOs, but also by the constraint from this second population of objects, which are unstable in the presence of large-$a$, low-$e$ Planet Nine orbits (and are additionally unstable in the presence of no Planet Nine at all). These two populations explain the unintuitive structure of the posterior probability distribution given in Figure \ref{fig:stab_lil}. 

Even though we require a Planet Nine iteration that does not cause the aligned TNOs to migrate significantly (more than 100 AU) over secular timescales, we expect there is a further population of objects that do migrate in the presence of Planet Nine. Indeed, some populations of objects in our solar system can be explained by this process. \cite{bbincl} uses the nominal Planet Nine orbit from \citet{bb16} to explain the existence of a population of highly inclined KBOs with $a\le$100: these objects can be explained as a migratory end-state of objects that started as members of the extreme TNO populations. 
The fact that some TNOs appear to experience migratory instabilities under the influence of Planet Nine is not inconsistent with our assumptions, but it does engender further questions. Clearly, some degree of migration is acceptable and will not alter the orbital alignment of the TNOs - the 100 AU threshold we use allows significant movement in semi-major axis without declaring objects dynamically unstable. However, the fact that objects can migrate in semi-major axis but remain confined to a comparably narrow range (disallowing above 100 AU of migration does not allow objects to move into entirely different object populations as seen in \citealt{bbincl}) begs the question: how are objects moving when they migrate in what we have defined as a dynamically \emph{stable} way? The answer to this question will be addressed in the following section.

\section{Proximity of the TNO Orbits to Resonances with Planet Nine}
\label{sec:resonance}

In the previous section, we identified that for some TNOs, migration in semi-major axis is an outcome for a significant fraction of integrations. In this section, we delve deeper into the nature of that migration and consider the question of resonance between the TNOs and Planet Nine.
Specifically, if the TNOs migrate in semi-major axis, they are likely to pass through the locations of mean-motion commensurabilities with Planet Nine, and one might expect the TNOs to fall into the stable configurations afforded by resonances, and stop migrating. 

The fact that proximity to resonance boosts dynamical stability can be directly applied to the Planet Nine  - TNO system. In particular, \citet{renu} considered the possibility that Planet Nine should be at an orbital radius that would allow it to be closest to low-order resonances for several TNOs, based on the currently measured orbits of those TNOs. The benefit to this configuration is that Planet Nine could stabilize the orbits of TNOs such as 2007 TG$_{422}$, which might otherwise migrate in semi-major axis. 
Similarly, \citet{sarah} used numerical N-body simulations to determine the best orbit of Planet Nine by testing which locations close to resonances allow the observed alignment of the TNOs. \citet{sarah} found the best semi-major axis of Planet Nine to be $a_{9} \sim$  654 AU (with a $e_{9} \sim$ 0.45), and \citet{renu} found the best-fit orbit to have a semi-major axis of $a_{9}\sim$ 665 AU. 

\begin{figure*} 
   \centering
   \includegraphics[width=7in]{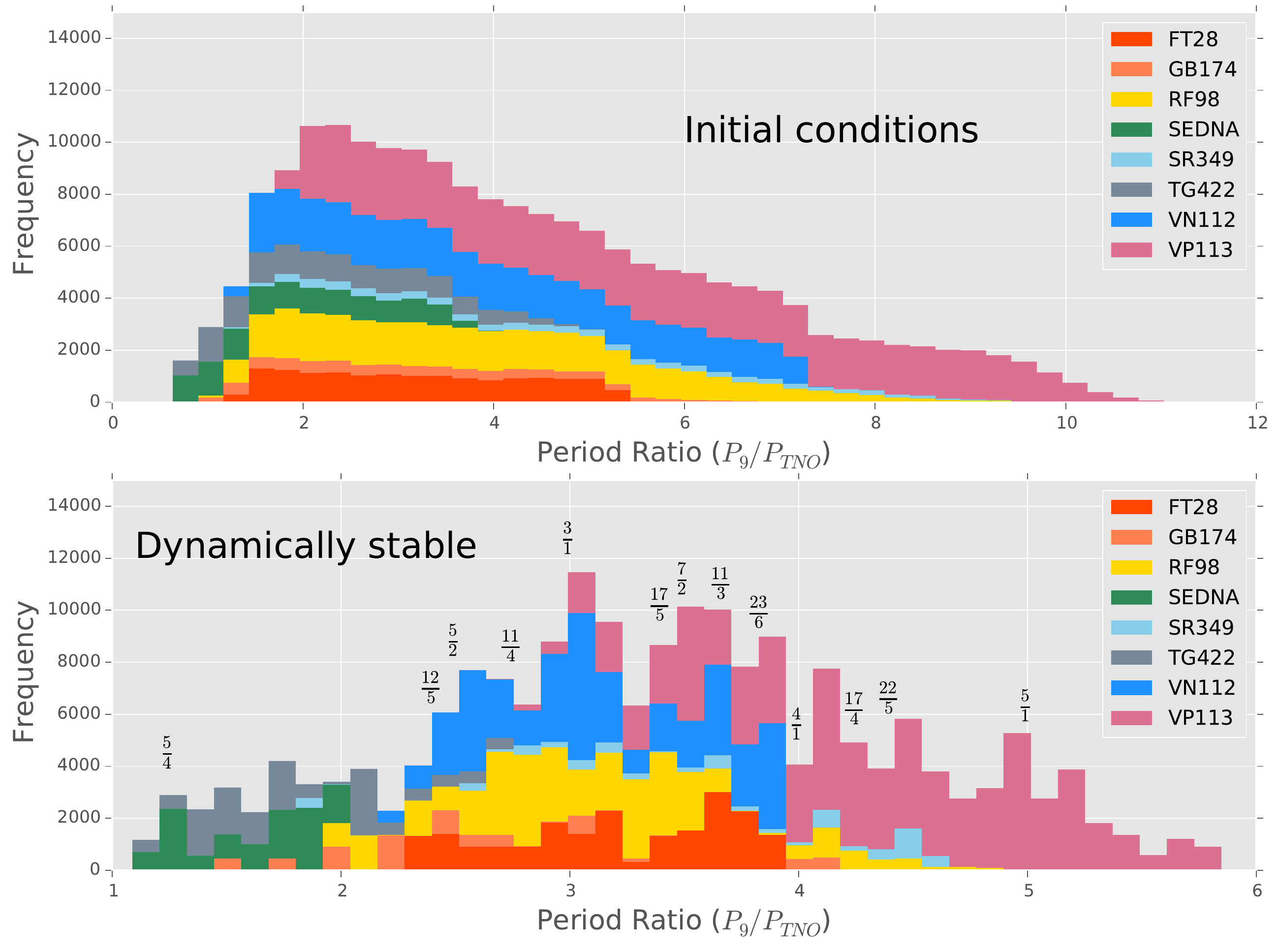} 
   \caption{(top panel) A sample of period ratios $P_{9} / P_{TNO}$, drawn from the initial conditions of the N-body simulations. This histogram shows the period ratio distribution that we would expect to see from the simulations if no period ratio were more dynamically stable than any other period ratio. This set serves as the control, and does not have the condition of dynamical stability imposed.  (bottom panel) Period ratios $P_{9} / P_{TNO}$ for the dynamically stable integrations at time-steps of one million years, demonstrating a peaked distribution, as some period ratios are preferred to others. In both panels, colors correspond to the values for each TNO, and the area of histograms was normalized. }
   \label{fig:res}
\end{figure*}

Clearly, resonance is an important aspect of the Planet Nine problem, as suggested by \citet{bb16}, \citet{beust}, \citet{renu}, and \citet{sarah}. In Figure \ref{fig:res}, we plot two histograms of the orbital period ratio ($P_{9} / P_{TNO}$) for each of the eight TNOs in our sample. In the top panel, we plot random draws from our initial conditions that were used to initialize the N-body simulations described in Section \ref{sec:num_methods}. The total number of draws in the top histogram was chosen to match the number of dynamically stable time-steps for each individual TNO, but the draws are from a raw distribution of the ratio $P_{9} / P_{TNO}$ for all numerical trials, including those that are not dynamically stable.
This top panel does not include the effect of certain ratios becoming more common due to gravitational interactions. In the bottom panel, we plot the dynamically stable trials for each TNO, with the period ratio sampled every million years.
The sharp peaks in this second histogram demonstrate clear overabundances of particular period ratios $P_9/P_{TNO}$. These values occur near resonances --- in the trials where dynamical stability is found --- and is markedly different from the continuum of period ratios shown in the upper panel. Keep in mind that the initial conditions of our simulations were chosen independently of resonance locations, unlike the set of simulations in \citet{sarah}. As a result, the behavior shown in the figure indicates that systems found to be dynamically stable in the simulations also show a clear preference for near-resonant locations. 

The results of this work show an important departure from the assumptions used previously \citep{renu,sarah}. This earlier work assumes that the TNOs remain in a single resonant configuration over the lifetime of the Solar System. For example, Sedna might live in either the 9:8 or 6:5 mean motion commensurability, depending on the semi-major axis of Planet Nine, but it is considered to remain in a single resonance. In this approximation, the orbits of the extreme TNOs that we observe today are consistent with their past orbits. This assumption allows for the estimate found in \citet{renu} to be carried out: if the TNOs reside in the same orbits over their entire lifetimes, and if they must be near resonance, then their current orbital properties can be used to compute the expected orbital elements of Planet Nine. However, our simulations show that TNOs do not always remain in the same orbits: Although they often remain near some resonance with Planet Nine, the TNOs change orbits and hence change resonances. 

\begin{figure} 
   \centering
   \includegraphics[width=3.4in]{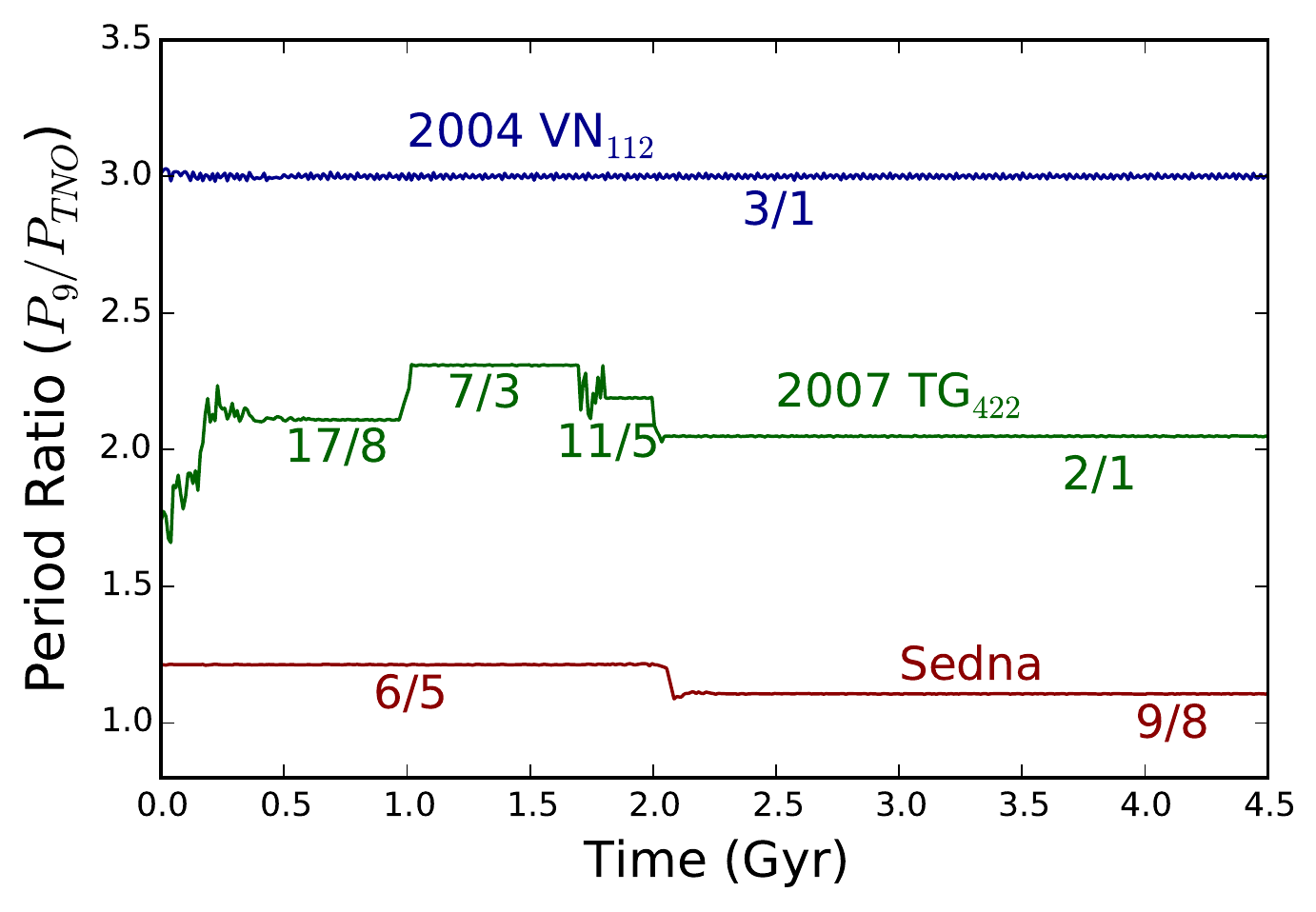} 
   \caption{For three integrations (top curve: 2004 VN$_{112}$; middle curve: 2007 TG$_{422}$; bottom curve: Sedna), we plot the period ratio $P_{9} / P_{TNO}$ as function of simulation time. The numbers denote resonances within $\Delta a_{\rm{res}}$ of the semi-major axis of each TNO, at each time-step. The top time series, showing the evolution of 2004 VN$_{112}$ during a typical integration, demonstrates how an object might remain in a single mean-motion commensurability for the entire lifetime of the solar system. The middle series, for 2007 TG$_{422}$, shows the behavior we call `resonance hopping,' where a TNO attains multiple mean-motion commensurabilities over the course of the simulation. The bottom series shows a less extreme version of this `resonance hopping,' where Sedna switches between two commensurabilities during the solar system lifetime. In all cases, the period of Planet Nine does not change over the course of the simulation: it is the motion of the TNO that leads to changing values of $P_{9}/P_{TNO}$. }
   \label{fig:example}
\end{figure}

In Figure \ref{fig:example}, we plot time series for 2004 VN$_{112}$ (top), 2007 TG$_{422}$ (middle), and Sedna (bottom) to demonstrate the potential resonant outcomes. The behavior of 2004 VN$_{112}$ is consistent with the assumption made by \citet{renu} and \citet{sarah}: that TNOs would live in a single resonance for the age of the solar system. The other two TNOs plotted in Figure \ref{fig:example} show behavior we call `resonance hopping,' where a TNO attains multiple mean-motion commensurabilities over the course of the simulation. When liberated from one mean-motion commensurability, both TNOs are captured into another resonance instead of being ejected from the system entirely. Planet Nine's semi-major axis did not change over the course of a single simulation, so all the change in $P_{9} / P_{TNO}$ within a single integration is due to migration by the TNO. 

As illustrated in Figure \ref{fig:stab_map_giant}, each TNO has a different dynamically stability map, preferring different regions of Planet Nine's potential parameter space. Similarly, when we take the subset of dynamically stable integrations for each TNO, each TNO has a different behavior relating to resonance. 
In this work, we take the definition of being `in' resonance to be living close to a resonant period ratio; as with the exoplanetary systems near resonance, the boost in dynamically stability provided by proximity to resonance applies even when systems are not in a perfect resonance. 
\citet{renu} and \citet{sarah} use a criterion for the proximity to resonance that is close enough to afford such benefits:
\be
\Delta a_{\rm{res}} \approx 0.007 a_{\rm{TNO}} \mid \frac{m_{9} a_{\rm{TNO}} \mathcal{A}}{3 M_{\odot} a_{9}} \mid^{1/2}
\label{eq:res}
\ee 
when $\Delta a_{\rm{res}}$ is the width, in AU, of the band of space close enough to a resonance to count as being `near' said resonance, subscript $TNO$ denotes the TNO's semi-major axis $a$ and the subscript 9 denotes Planet Nine's semi-major axis $a_{9}$ and mass $m_{9}$. $\mathcal{A}$ is a unitless coefficient.
We choose to use $\mathcal{A}=3$ as done in \citet{renu} and \citet{sarah}. 
The numerical value of $\Delta a_{\rm{res}}$ tends to be close to 5-10 AU for the TNOs in our sample, meaning that the simulation results allow us to determine the nearest resonance and identify TNOs that `hop' between resonances. Figure \ref{fig:example} shows three examples from our set of simulations: 2004 VN$_{112}$ does not change resonances during the integration, 2007 TG$_{422}$ changes resonances three times, and Sedna changes resonance once. For two objects in our sample (2004 VN$_{112}$ and 2012 VP$_{113}$), we see no hopping behavior.

\begin{figure} 
   \centering
   \includegraphics[width=3.4in]{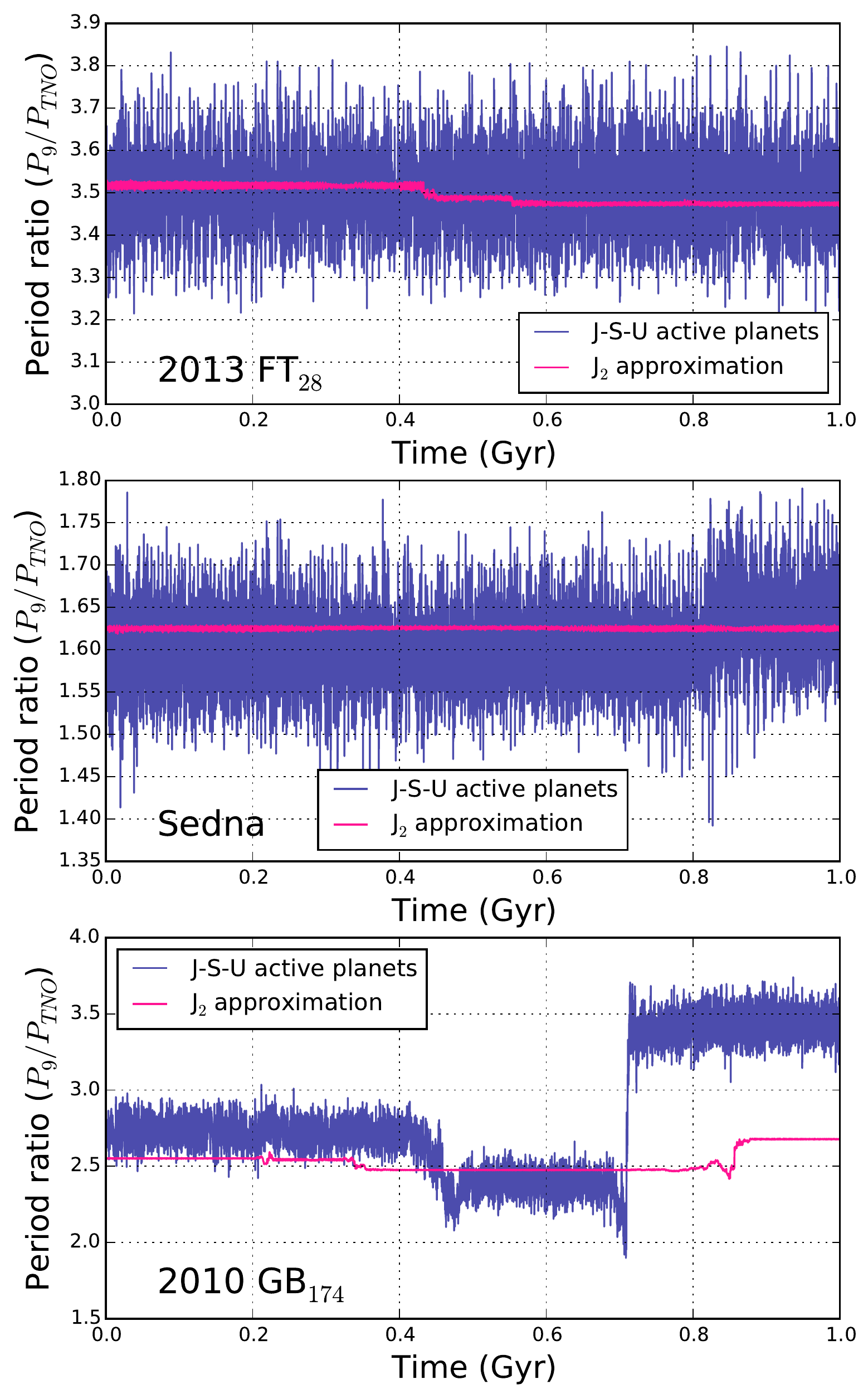} 
   \caption{A check of our integrations (which used solar J$_{2}$ in place of the inner three giant planets) and integrations run with active Jupiter, Saturn, and Uranus (JSU) show that although both sets exhibit the same hopping behavior, the J$_{2}$ approximation underestimates the noise in period ratio and overestimates the degree to which the resonances can be differentiated. }
   \label{fig:res_hop}
\end{figure}

However, it is important to note that the simulations used to construct Figure \ref{fig:res} and the curves in Figure \ref{fig:example} are the set run for this paper, which uses the quadrupole moment of the central part of the system ($J_{2}$) to replace the active motions of Jupiter, Saturn and Uranus (JSU). This approximation allows the simulations in this work to be completed in a total of roughly 100,000 total CPU hours, rather than the nearly 2 million CPU hours that would be required to run the full integrations with all active giant planets. This approximation is appropriate for the tests of dynamical stability and alignment considered thus far in this work, but when considering the question of resonance, some discrepancies arise. In Figure \ref{fig:res_hop}, we present a comparison between a few test simulations run with the $J_{2}$ approximation and the JSU case of active particles for Jupiter, Saturn, and Uranus. The test simulations were run for 1 Gyr each, with a single realization of Planet Nine ($a$ = 700 AU, $e$ = 0.5) and otherwise identical to the cases run in the previous set of simulations. For the JSU set of simulations, we lowered the time-step to 20 days.

When active JSU particles are included in the integrations, two major differences occur as compared to the J$_{2}$ approximation: (1) the period ratios are not as tightly confined, experiencing a larger degree of scatter even while living in a single apparent resonance; (2) the number of times objects hop between resonances can be both increased (due to repeated accelerations from Uranus) and decreased (as shown in the top panel of Figure \ref{fig:res_hop}, hops we resolve in the J$_{2}$ case are un-physical in the JSU case).

Regarding point (1), the second panel of Figure \ref{fig:res_hop} illustrates a case where in both the J$_{2}$ and JSU cases, a TNO lives close to the 8/5 mean motion resonance. In the J$_{2}$ case, the TNO stays within 0.3\% of its average period ratio. In the JSU case, the TNO stays within 5\% of its averege period ratio (this value is larger than in the J$_{2}$ case due to the inclusion of accelerations from the inner three giant planets). Due to this complication, we have chosen the bin size in Figure \ref{fig:res} to be commensurate with the period ratio confinement experienced by TNOs in the JSU case, which is typically around 8\%. 

Regarding point (2), it is unclear without running a large suite of JSU integrations how the hopping frequency changes with the inclusion of the giant planets. For this reason, we do not present in this work detailed results of the resonance hopping in our simulations. In order to accurately assess this behavior, a complete set of simulations with active giants planets (JSU) should be carried out. 

Our simulations show that dynamically stable integrations of the TNOs tend to attain mean-motion commensurabilities. Depending on the TNO, it may attain a single resonant location and stay there, or it may hop between resonant locations. This `resonance hopping' is an important effect, and for TNOs that exhibit this behavior, their past semi-major axis may be different than the current values. The numerical computation of the specifics of this behavior should be a fruitful avenue for future work.

%
%
%
%
%
%
%
%
%

%
%
%
%
%
%
%
%

%
%
%
%
%
%
%
%
%
%
%
%
%
%
%
%
%
%
\section{Conclusions}
\label{sec:conclusion}
In this work, we have evaluated the dynamical stability and orbital alignment of eight TNOs (Sedna, 2004 VN$_{112}$, 2007 TG$_{422}$, 2010 GB$_{174}$, 2012 VP$_{113}$, 2013 RF$_{98}$, 2013 FT$_{28}$, and 2014 SR$_{349}$) in the presence of a Monte Carlo assortment of Planet Nine realizations with varying semi-major axis and eccentricity.
We used the results to predict the most probable ($a,e$) of Planet Nine by deriving the posterior probability distributions for Planet Nine's orbital elements ($a,e$). The distribution based on dynamical stability considerations for the TNOs is presented in Figure \ref{fig:stab_lil}. We have also constructed an analogous probability distribution based on the requirement that the orbits of the TNOs remain aligned. Both of these posterior distributions demonstrate that the preferred orbits for Planet Nine have intermediate values of eccentricity ($0.3<e<0.5$) and semi-major axis ($650<a<900$ AU), as shown in Figure \ref{fig:test}. Moreover, these  values are roughly consistent with the regime suggested in \citet{bb162}, which constructed its probability map using clustering arguments only. Our stability posterior and that from \citet{bb162} were constructed based on different fundamental orbital properties (dynamical stability and secular evolution patterns, respectively). Despite this significant difference in construction, the two results are consistent, in that they both prefer non-zero eccentricities and a similar range in semi-major axis for Planet Nine. The comparison between our results and those of \citet{bb162} is shown in Figure \ref{fig:test}. Notably, similar dynamical stability arguments in \citet{sarah} produce a best-fit Planet Nine of 654 AU and 0.45 eccentricity, which is consistent with the overlap region between the results of this work and of \citet{bb162}. 

Using numerical N-body simulations, we also demonstrated that 2007 TG$_{422}$ and 2013 RF$_{98}$, while found in the past to be dynamically unstable in the presence of Neptune alone, can attain dynamically stable states in the presence of Planet Nine. Our simulation results support the prediction of \citet{ST_tnos} that since 2007 TG$_{422}$ and 2013 RF$_{98}$ exhibit the same orbital clustering as the dynamically stable TNOs, Planet Nine likely dominates over Neptune interactions.
In addition, we find that different TNOs exhibit very different stability maps, with some objects (such as Sedna and 2012 VP$_{113}$) contributing relatively little unique information to the stability posterior and others (such as 2007 TG$_{422}$ and 2013 RF$_{98}$) exhibiting unintuitive preference against large-$a$, low-$e$ orbits of Planet Nine. 
These two categories suggest that there may be two dynamical classes of objects in this TNO sample, which interact differently with Planet Nine. However, we have considered in this work only a small ($N=8$) number of objects that fit into the desired high-$a$, trans-Neptunian-$q$, apsidally aligned category identified in \citet{bb16}. The discovery of additional objects in this population (expected in the near future) will allow for a more robust test of this two-population hypothesis. 

We have also evaluated the different dynamical outcomes for these extreme TNOs in the presence of Planet Nine. The objects that are dynamically unstable in the presence of large-$a$, low-$e$ orbits of Planet Nine (2007 TG$_{422}$ and 2013 RF$_{98}$) tend to experience migration rather than violent collisions or ejections as their main outcome in dynamically unstable cases. These objects are also dynamically unstable in the presence of only Neptune and the other giant planets, i.e., in the absence of Planet Nine. In Table \ref{tab:stab}, we present the difference $\bar{\delta t}$ in average dynamical lifetime between the case where migration is considered to be a dynamical instability mechanism and when it is not. Table \ref{tab:tabtwo} presents the relative occurrence rates for each type of outcome. For cases where the TNOs are not stable over the lifetime of the Solar System, the fraction of trials that lose objects to migration (with $a>100$ AU), close encounters with giant planets, and ejection from the system are roughly comparable. A small minority of the simulations end with accretion onto the Sun (less than 1\%). 

Next, we suggest a generalized description for the interactions between the TNOs and Planet Nine. We propose that the paradigm is neither that (1) mean motion resonance is unimportant, nor (2) TNOs reside in a single resonance with Planet Nine for the age of the solar system. Instead, while some TNOs (such as 2004 VN$_{112}$ and 2012 VP$_{113}$) can sometimes live in a single resonance for solar system lifetimes, others exhibit a behavior that we call `resonance hopping.' This term means that the TNO is near-continually in close proximity to a mean motion resonance (Figure \ref{fig:res}), but it is not necessarily near the \emph{same} resonance for the age of the Solar System (Figure \ref{fig:example}). Instead, the TNOs can transition between closely-spaced resonances, often those described by relatively large integer ratios. The long-term effect of this process is that the orbital anti-alignment caused by Planet Nine is able to persist, but the TNO is protected against small kicks in energy provided by interactions with Neptune. In this paradigm, an interaction with Neptune might lead to the movement of a TNO into a new resonance, but not to its ejection from the solar system (see also \citealt{renu,sarah}). A useful avenue for future work would be the full numerical computation of this effect, as the $J_{2}$ approximation used in this work precludes an accurate calculation of the true frequency of resonance hopping. 

Another important avenue for future work exists in our prediction that the eight TNOs considered in this work populate two distinct dynamical classes: first, a class of objects such as 2012 VP$_{113}$ or Sedna, which are more dynamically stable in the presence of large $a$, low $e$ realizations of Planet Nine (which reduce closely to the case of having no ninth planet); second, a class of objects that are dynamically unstable in the presence of only Neptune and no Planet Nine (and also in the case of a high-$a$, low-$e$ Planet Nine). This second class of objects may have higher $e$ and $a$ than the first, and require a stabilizing influence in the form of an eccentric Planet Nine to prevent destabilizing interactions with Neptune. To truly understand if this is a valid division, it is hoped that a large number of high-$a$, trans-Neptunian-$q$, apsidally aligned TNOs will be discovered in the next few years. The classification of these objects, and exploration of the mechanism by which Planet Nine may stabilize their orbits, should be explored in the future.

%
%
%
%
%
%
%
%
%
%
%
%
%

\acknowledgements

We thank Konstantin Batygin for a thorough review of the manuscript and many useful suggestions. We thank Andrew Vanderburg, Sarah Millholland, Jaehan Bae, Marina Kounkel, and Ellen Price for useful conversations. We thank Michael Dieterle and Clara Eng for helpful suggestions on how to effectively visualize the simulation results. We also thank Gary Bernstein for reviewing the manuscript and suggesting several improvements. Finally, we thank the referee for many useful comments that improved the paper. This work was supported by NSF Grant AST-1515015. J.C.B and S.J.H. are also supported by the NSF Graduate Research Fellowship Grant No. DGE 1256260. The computations for this work used the Extreme Science and Engineering Discovery Environment (XSEDE), which is supported by National Science Foundation grant number ACI-1053575. This research was done using resources provided by the Open Science Grid, which is supported by the National Science Foundation and the U.S. Department of Energy's Office of Science.

\bibliographystyle{apj}
\bibliography{refs}

\vskip 0.50truein

\begin{appendices}
\section{Appendix: Precession Equations of Motion for TNOs}  
\label{app1}

To consider the secular motion of the TNOs in the presence of Planet Nine, we can treat the problem with a coplanar approximation, as done in \citet{bb16}. It is important to note that this is a very rough approximation, as the TNOs and Planet Nine are expected to be inclined relative to the inner solar system and each other. 

The Hamiltonian for this system, as used in \cite{bb16}, is: 
\be
\mathcal{H} = -\frac{1}{4} \frac{GM}{a} (1-e^{2})^{-3/2} \sum^{4}_{i=1}\left( \frac{m_{i}a_{i}^{2}}{M a^{2}}\right) -\frac{Gm_{9}}{a_{9}}
\bigg[ \frac{1}{4} \left(\frac{a}{a_{9}}\right)^{2} \frac{1+ 3e^{2} / 2}{(1-e_{9}^{2})^{3/2}}
-\frac{15}{16} \left(\frac{a}{a_{9}}\right)^{3} e e_{9} \frac{1+ 3e^{2} / 4}{(1-e_{9}^{2})^{5/2}} \cos{(\omega_{9}-\omega)}
\bigg]
\label{hamiltonian}
\ee
when $M$ is the mass of the central body, the subscript $9$ denotes the properties of Planet Nine, the subscript $i$ within the summation in the first term denotes the four giant planets, and a lack of subscript denotes the TNO for which the Hamiltonian is being written. 

Using canonical variable $\epsilon = \sqrt{1-e^{2}}$, we can find the equation of motion for the argument of perihelion $\omega$ by taking the derivative of Equation \ref{hamiltonian}, such that $d \omega / dt \propto d \mathcal{H} / d \epsilon$. The result of this yields
\be
\frac{d\omega}{dt} \propto \bigg[ \frac{3}{4} \sum^{4}_{i=1} m_{i}a_{i}^{2} (e^{2} -1)^{-2} a^{-2} -\frac{15}{16} m_{9} \frac{e_{9}}{e} \left( \frac{a}{a_{9}}\right)^{4} \frac{\sqrt{1-e^{2}}}{(1-e_{9}^{2})^{5/2}} (1- 9e^{2}/4) \cos(\omega - \omega_{9}) -\frac{3}{4} m_{9} \left( \frac{a}{a_{9}}\right)^{3} \frac{\sqrt{1-e^{2}}}{(1-e_{9}^{2})^{3/2}}\bigg],
\label{eom}
\ee
when leading constants have been dropped, since in this work we care only about the relative contributions of the terms of the equation of motion. 
The first term of Equation \ref{eom} represents the apsidal precession of a TNO with orbital elements ($a,e,\omega$) that is caused by the inner solar system (where the outer four giant planets are treated as a solar oblateness and the terrestrial planets ignored).
The latter two terms of Equation \ref{eom} include dependences on the mass of Planet Nine ($m_{9}$), eccentricity of Planet Nine ($e_{9}$), semi-major axis of Planet Nine ($a_{9}$), and argument of perihelion of Planet Nine ($e_{9}$). These two terms represent the apsidal precession of the TNO due to Planet Nine's influence. 

Alone, the influence of Planet Nine \emph{or} of the inner solar system would lead to precession of each TNO's $\omega$. Taken together, the two precession terms can either boost precession rates or slow them. With the proper choice of orbital elements for Planet Nine, $d\omega /dt$ can be set to be zero, leading to no net precession relative to the Katti-Range vector of Planet Nine's orbit. Such a situation could result in a selection of TNOs exhibiting orbits that remain in roughly the same regime of parameter space over time, potentially leading to alignment like that observed in the TNOs in our solar system. 


For completeness, we note that Equation \ref{eom} can also be derived from the disturbing function using Lagrange's planetary equations. 
We use the disturbing function as formulated in \citet{mardling}, which is written to work in the dimensions of energy. The disturbing function can be written as 
\be
\mathcal{R} = -\frac{1}{4} \frac{G M m}{a} (1-e^{2})^{-3/2} \sum^{4}_{i=1}\left( \frac{m_{i}a_{i}^{2}}{M a^{2}}\right) + \frac{G m_{9}m}{a_{9}} \left[\frac{1}{4} \left(\frac{a}{a_{9}} \right)^{2}  \frac{1 + 3/2 e^{2} }{(1-e_{9}^{2})^{3/2}}  -\frac{15\ e \ e_{9}}{16}  \left( \frac{a}{a_{9}}\right)^{3} \frac{1+ 3 e^{2}/4}{(1-e_{9}^{2})^{5/2}} \cos(\omega - \omega_{9}) \bigg]\right],
\label{disturb}
\ee
This equation has been simplified under the assumption $m << m_{p} << M$. We can use Lagrange's planetary equations to find the equation of motion analogous to that in Equation \ref{eom}. Specifically:
\be
\frac{d\omega}{dt} = \frac{\epsilon}{m \nu a^{2} e} \frac{d\mathcal{R}}{de}
\label{lagrnage}
\ee
when $\nu$ is the orbital frequency of the TNO. Substituting Equation \ref{disturb} into the relevant Lagrange equation (Equation \ref{lagrnage}) yields the full equation of motion:
\be
\frac{d\omega}{dt} = \frac{m_{9}\nu}{M} \bigg[ \frac{3}{4} \sum^{4}_{i=1} m_{i}a_{i}^{2} (e^{2} -1)^{-2} a^{-2} -\frac{15}{16} m_{9} \frac{e_{9}}{e} \left( \frac{a}{a_{9}}\right)^{4} \frac{\sqrt{1-e^{2}}}{(1-e_{9}^{2})^{5/2}} (1- 9 e^{2}/4) \cos(\omega - \omega_{9}) -\frac{3}{4} m_{9} \left( \frac{a}{a_{9}}\right)^{3} \frac{\sqrt{1-e^{2}}}{(1-e_{9}^{2})^{3/2}}\bigg],
\label{eom2}
\ee 
which is equivalent to the result presented in Equation \ref{eom}. 

It is very important to note that the preceding derivation makes several major approximations:
\begin{itemize}
\item We assume a coplanar system and neglect orbital inclination of all bodies,
\item In the construction of the Hamiltonian and the disturbing function, we ignore all short-order, resonant terms\footnote{and as shown in Section \ref{sec:resonance}, this is probably dangerous to do}
\end{itemize} 
For all of these reasons, this analytic result should be treated as approximate. To do the problem properly, it is important to use numerical N-body simulations. 

\section{Appendix: Effects of allowing orbital elements to vary in N-body simulations}
\label{app2}
In the set of N-body simulations that we describe in Section \ref{sec:num_methods} and use to construct the posterior probability distribution given in Section \ref{sec:bayes}, we use a population of Planet Nine realizations with fixed inclination $i$ = 30 degrees, argument of perihelion $\omega$ = 150 degrees, and longitude of the ascending node $\Omega$ = 113 degrees. This was done because the amount of uncertainty in each of these measurements would require a computationally unfeasible  number of integrations to well-sample the parameter space. Instead, we chose the approximate best values for each angle, as reported and used in prior literature. 

However, we can recreate the probability posterior presented in Figure \ref{fig:stab_lil} for an additional set of 1500 N-body integrations, while allowing these orbital angles to vary, and examine the amount of difference between this new posterior and the old one as a first test. To do this, we ran 1500 more numerical N-body integrations with the same numerical properties as our other set (hybrid symplectic and Bulirsch-Stoer (B-S) integrator in \texttt{Mercury6} \citep{m6}, conserving energy to 1 part in $10^{10}$, replacing the three inner giant planets with a solar J2, including each TNO with orbital elements drawn from observational constraints).
In this set of integrations, we sampled $a_{9}$ from a uniform range between (400, 1200) AU, and $e_{9}$ between (0,1). However, instead of fixing the orbital angles, we sampled from normal distributions centered on the \citet{bb16} estimates ($i_{9}$ = 30 degrees,  $\omega_{9}$ = 150 degrees, $\Omega_{9}$ = 113 degrees) with widths of 30 degrees in each case. 
Then, we integrated each of the 1500 realizations forward for 4.5 Gyr.

\begin{figure} 
   \centering
   \includegraphics[width=3.4in]{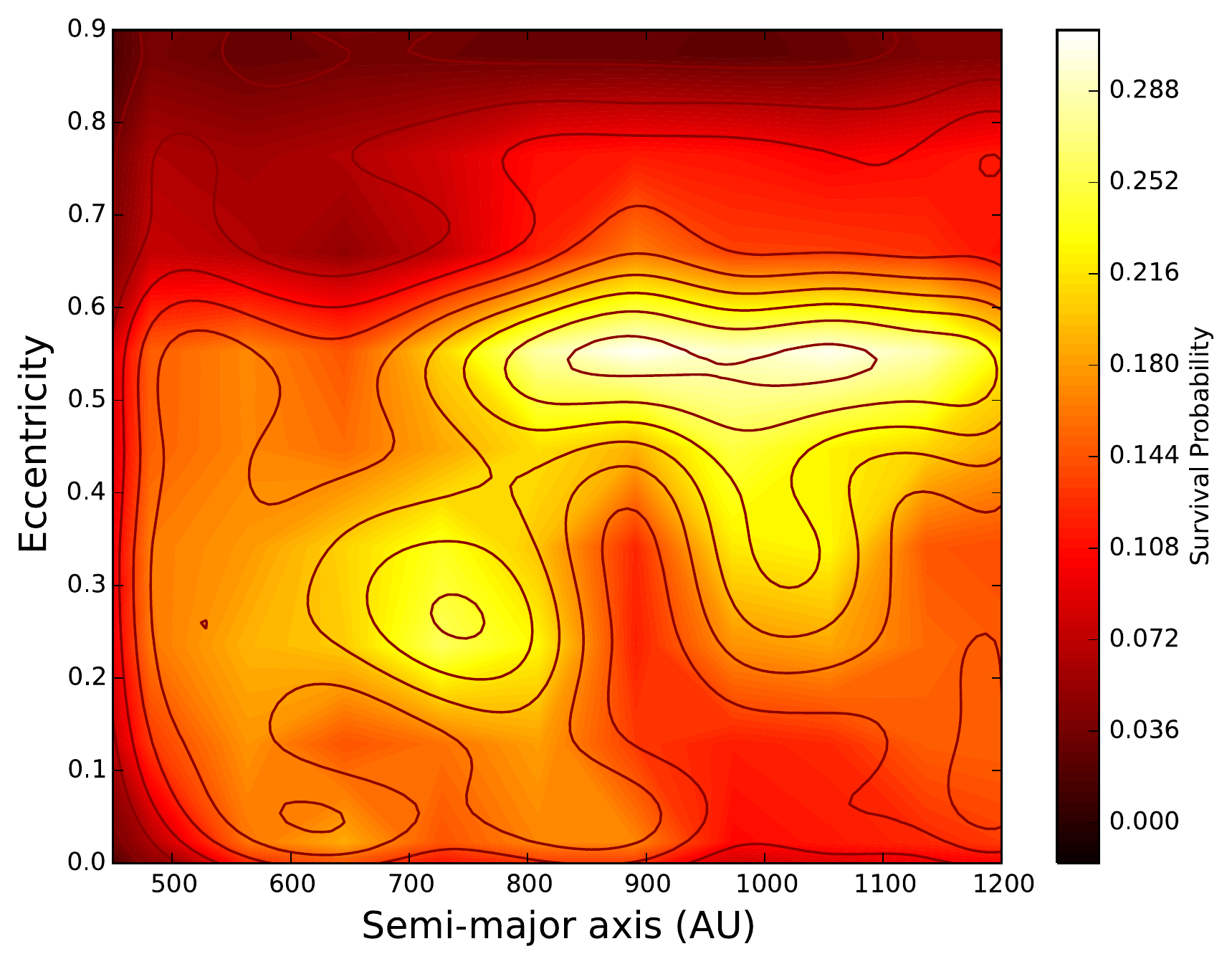} 
   \caption{The overall stability posterior for the semi-major axis and eccentricity of Planet Nine, when Planet Nine's orbital angles were allowed to vary over a normal distribution centered on the best-guess values. This posterior was constructed by taking a summation of the posteriors for each individual object, including the six objects used in \cite{bb16} and the two new high-$a$, low-$e$ objects from \citet{ST_tnos}. As compared to Figure \ref{fig:stab_lil}, which did not allow the orbital angles to vary, the survival probability is lower when orbital angles are allowed to vary, suggesting that the best-guess values leading to more probable alignment also lead to more dynamically stable configurations.}
   \label{fig:stab_lil2}
\end{figure}

The results of this new, second suite of integrations is presented in Figure \ref{fig:stab_lil2}, which can be directly compared to Figure \ref{fig:stab_lil} (which was constructed with our original set of simulations). Comparing the two figures presents three major conclusions: [1] The main parameter space preferred in each set is similar, with high eccentricities being less preferred and the range from 0.3-0.5 eccentricity, 600-1000 AU being good in both sets. [2] The overall survival probabilities are lower for the case where Planet Nine's orbital angles are allowed to vary, which indicates that altering these values too much decreases dynamical stability of the TNOs overall. [3] The posterior presented in Figure \ref{fig:stab_lil2}, with varying orbital angles, has less variation between peaks and valleys, which is the danger of adding additional layers of variation (which we did by allowing three orbital angles to vary).

The reason that we do not use this set of simulations that allow initial orbital angles of Planet Nine to vary as the main set in this paper is that the parameter space is not well-sampled. Additionally, the three additional free parameters bring with them three additional sources of variations. The robust exploration of this parameter space is outside the scope of this work. 
\end{appendices}

\end{document}